\documentclass[11pt]{article}
\usepackage[utf8]{inputenc}

\title{Configuration Balancing for Stochastic Requests}
\author{
Franziska Eberle \\ London School of Economics \\ f.eberle@lse.ac.uk
\and Anupam Gupta \\ Carnegie Mellon \\ anupamg@cs.cmu.edu 
\and Nicole Megow \\ University of Bremen \\ nicole.megow@uni-bremen.de
\and Benjamin Moseley \\ Carnegie Mellon\\ moseleyb@andrew.cmu.edu
\and Rudy Zhou \\ Carnegie Mellon \\ rbz@andrew.cmu.edu}
\date{\today}

\usepackage{amsmath,amsthm,nicefrac,mathtools}
\usepackage{amsfonts, amstext}

\usepackage[margin=1in,letterpaper]{geometry}
\usepackage[font={small,it}]{caption}

\usepackage{setspace}
\usepackage{comment}

\usepackage{enumitem}
\usepackage[breaklinks]{hyperref}
\hypersetup{colorlinks=true,%
            citebordercolor={.6 .6 .6},linkbordercolor={.6 .6 .6},%
citecolor=blue!50!black,urlcolor=black,linkcolor=blue!50!black,pagecolor=black}

\usepackage[nameinlink]{cleveref}
\Crefname{algocf}{Algorithm}{Algorithms}
\crefname{algocfline}{line}{lines}
\Crefname{invariant}{Invariant}{Invariants}
\Crefname{claim}{Claim}{Claims}
\Crefname{subclaim}{Subclaim}{Subclaims}
\Crefname{remark}{Remark}{Remarks}

\usepackage{epsfig}
\usepackage{amsthm,amssymb}

\usepackage{mathrsfs}
\usepackage{xspace}
\usepackage{soul}
\usepackage{latexsym}
\usepackage{bbm}

\usepackage{framed}

\usepackage[dvipsnames]{xcolor}
\definecolor{DarkGray}{rgb}{0.66, 0.66, 0.66}
\definecolor{DarkPowderBlue}{rgb}{0.0, 0.2, 0.6}
\definecolor{fluorescentyellow}{rgb}{0.8, 1.0, 0.0}

\usepackage[ruled,vlined]{algorithm2e}
\SetKwFor{Upon}{upon}{do}{end}

\SetCommentSty{mycommfont}

\usepackage{thmtools,thm-restate}

\allowdisplaybreaks

\newcommand{\alert}[1]{{\color{red}#1}}

\usepackage[textsize=tiny,textwidth=2cm,color=Red]{todonotes}

\newcommand{\nic}[1]{\todo[color=LimeGreen]{N: #1}}
\newcommand{\rud}[1]{\todo[color=Salmon]{R: #1}}

\newcommand{\addref}{\todo[color=Red]{Add reference}}
\newcommand{\add}[1]{\todo[color=Red, inline, size=\normalfont]{Add #1}}

\newcounter{note}[section]

\sethlcolor{fluorescentyellow}

\newcommand{\initOneLiners}{%
    \setlength{\itemsep}{0pt}
    \setlength{\parsep }{0pt}
    \setlength{\topsep }{0pt}
}

\newtheorem{theorem}{Theorem}[section]

\newtheorem{lem}[theorem]{Lemma}

\theoremstyle{definition}
\newtheorem{definition}[theorem]{Definition}
\newtheorem{example}[theorem]{Example}

\theoremstyle{remark}

\newcommand{\EE}{\mathbb{E}}
\newcommand{\RR}{\mathbb{R}}

\newcommand{\Alg}{\textsc{Alg}}
\newcommand{\Opt}{\textsc{Opt}}

\newcommand{\lpconfig}{\ensuremath{\textsf{LP}_{\textsf{C}}}}

\newcommand{\lppath}{\ensuremath{\textsf{LP}_{\textsf{P}}}}
\newcommand{\lppathd}{\ensuremath{\textsf{D}_{\textsf{P}}}}

\newcommand{\PP}{\ensuremath{\mathbb{P}}}

\newcommand{\load}{L}
\newcommand{\one}{\ensuremath{\mathbbm{1}}}
\newcommand{\poly}{\operatorname{poly}}

\begin{document}

\maketitle
\begin{abstract}

  The configuration balancing problem with stochastic requests
  generalizes many well-studied resource allocation problems such as
  load balancing and virtual circuit routing. In it, we have $m$
  resources and $n$ requests. Each request has multiple possible
  \emph{configurations}, each of which increases the load of each
  resource by some amount. The goal is to select one configuration for
  each request to minimize the \emph{makespan}: the load of the
  most-loaded resource. In our work, we focus on a stochastic setting, where we only know the distribution for how each configuration increases the resource loads, learning the realized value only after a configuration is chosen. 

  \medskip
We develop both offline and online algorithms for configuration
balancing with stochastic requests. When the requests are known
offline, we give a non-adaptive policy for configuration balancing
with stochastic requests that $O(\frac{\log m}{\log \log
  m})$-approximates the optimal adaptive policy. In particular, this
closes the adaptivity gap for this problem as there is an
asymptotically matching lower bound even for the very special case of
load balancing on identical machines. When requests arrive online in a
list, we give a non-adaptive policy that is $O(\log m)$
competitive. Again, this result is asymptotically tight due to
information-theoretic lower bounds for very special cases (e.g., for load
balancing on unrelated machines). Finally, we show how to leverage
adaptivity in the special case of load balancing on \emph{related} machines to
obtain a constant-factor approximation offline and an $O(\log \log
m)$-approximation online. A crucial technical ingredient in all of our
results is a new structural characterization of the optimal adaptive policy that allows us to limit the correlations between its decisions.

\end{abstract}

\thispagestyle{empty}

\newpage
\setcounter{page}{1}

\section{Introduction}\label{sec_intro}

This paper considers the \emph{configuration balancing} problem.  In this problem there are~$m$ resources and~$n$ requests. Every request~$j$
has~$q_j$ possible configurations~$x_j(1), \ldots, x_j(q_j) \in
\RR_{\geq 0}^m$. We must choose one configuration~$c_j \in [q_j]$ per
request, which adds $x_j(c_j)$ to the load vector on the resources.
The goal is to 
minimize the makespan, i.e., the load of the most-loaded resource.  
Configuration balancing captures many natural resource allocation problems where requests 
	compete for a finite pool of resources and the task is to find a ``fair'' allocation in which no resource is over-burdened. 
 Two well-studied problems of this form arise in  scheduling and routing.
\begin{itemize}[nolistsep]
\item[(i)] In \emph{load balancing} a.k.a.~\emph{makespan
    minimization}, there are $m$ (unrelated) machines and $n$ jobs. Scheduling job $j$ on machine $i$ increases the load of~$i$ 
    by $p_{ij} \geq 0$. The goal is to schedule each job on
  some machine 
  to minimize the
  makespan (the load of the most-loaded machine).
\item[(ii)] In \emph{virtual circuit routing} or \emph{congestion
    minimization}, there is a directed graph $G=(V,E)$ on $m$ edges
  with edge capacities $c_e > 0$ for all $e \in E$, and $n$ requests,
  each request consisting of a source-sink pair $(s_j, t_j)$ in $G$
  and a demand $d_j \geq 0$. The goal is to route each request $j$
  from source to sink via some directed path, increasing the
  load/congestion of each edge $e$ on the path by
  $\nicefrac{d_j}{c_e}$. Again, the objective is to minimize the load
  of the most-loaded edge.
\end{itemize}

Configuration balancing captures both of these problems by taking
the~$m$ resources to be the~$m$ machines or edges, respectively;  
each configuration now corresponds to assigning a job to some machine
or routing a request along a particular source-sink
path.

Typically, job sizes or request demands are not
  known exactly when solving resource allocation problems in
  practice. This motivates the study of algorithms under uncertainty,
  where an algorithm must make decisions given only partial/uncertain
  information about the input. Uncertainty can be modeled in
  different ways. In exceptional cases, a {\em non-clairvoyant} algorithm that has 
  {\em no} knowledge about the loads of requests may perform surprisingly well; an example is Graham's greedy list scheduling for load balancing on identical machines~\cite{DBLP:journals/siamam/Graham69}.
However, in general, a non-clairvoyant algorithm 
cannot perform well. In this work, we consider a stochastic model, where the unknown input follows some known distribution but the actual realization is a priori unknown. Such a model is very natural when there is historical data available from which such distributions can be deduced.

In the {\em configuration balancing with stochastic requests}
problem, we assume that each configuration~$c$ of request~$j$ is a random
vector~$X_j(c)$ with known distribution such that the~$X_j(c)$'s are
independent across different requests~$j$. However, the actual
realized vector of a configuration~$c$ of request~$j$ is only observed
after \emph{irrevocably} selecting this particular configuration for
request~$j$. The objective is to minimize the expected maximum load
(i.e., the expected makespan)
\[
    \mathbb{E} \Big[\max_i \sum_{j=1}^n X_{ij}(c_j) \Big],  
\]
where $c_j$ is the configuration chosen for request $j$.

Further, we distinguish whether there is an additional dimension of uncertainty or not, namely the knowledge about the request set.
In the \emph{offline} setting, the {set of requests and the} distributions of the configurations of each request are known up-front, and they can be selected and assigned to the resources irrevocably in any order. 
In the \emph{online} setting, requests are not known in advance and they are revealed one-by-one (online-list model). The algorithm learns the stochastic information on configurations of a request upon its arrival, and must select one of them without knowledge of future arrivals. After a configuration is chosen irrevocably, the next request arrives. 

In general, we allow an algorithm to base the next decision on knowledge about the realized vectors of all previously selected request configurations. We call such policies \emph{adaptive}. Conversely, a {\em non-adaptive} policy is one that fixes the particular configuration chosen for a request without using any knowledge of the realized configuration vectors. 

The goal of this paper is to investigate the power of adaptive and non-adaptive policies for online and offline configuration balancing with stochastic requests. We quantify the performance of an algorithm by bounding the worst-case ratio of the achieved expected makespan and the minimal expected makespan achieved by an optimal offline adaptive policy. We say that an algorithm \Alg\ $\alpha$-\emph{approximates} an algorithm \Alg' if, for any input instance, the expected makespan of \Alg\ is at most a factor $\alpha$ larger than the expected makespan of \Alg'; we refer to $\alpha$  also as {\em approximation ratio}. For online algorithms, the term \emph{competitive ratio} refers to their approximation ratio.

\subsection{Our Results}

\bigskip \noindent {\bf Main result.\ \ }
As our first main result, we present non-adaptive algorithms for offline and online configuration balancing with stochastic requests.

\begin{theorem}\label{thm_config_off}\label{thm_config_balancing}
For configuration balancing with stochastic requests, when the configurations are given explicitly, there is an efficient polynomial-time randomized \emph{offline} algorithm that computes a non-adaptive policy that is a $\Theta\big(\frac{\log m}{\log \log m}\big)$-approximation and an \emph{online} algorithm that is a~$\Theta(\log m)$-approximation when comparing to the  optimal offline adaptive policy.
\end{theorem}
The offline analysis relies on a linear programming (LP) relaxation of configuration balancing. which has a known integrality gap of $\Theta\big(\frac{\log m}{\log \log m}\big)$, even for virtual circuit routing~\cite{LeightonRS98}, implying that the analysis is tight. 
In the online setting, our analysis employs a potential function to
greedily determine which configuration to choose for each request. In
particular, we generalize the idea by \cite{AspnesAFPW97} to the
setting of configuration balancing with stochastic requests and match a known lower bound for online load balancing on unrelated machines with deterministic jobs by~\cite{DBLP:journals/jal/AzarNR95}. 

If the configurations are not given explicitly, efficiently
  solving the problem requires us to be able to
  optimize over configurations in polynomial time.

\bigskip \noindent {\bf Applications.\ \ } These results would hold for both, load
balancing on unrelated machines and virtual circuit routing, if
we could guarantee that either the configurations are given
  explicitly or the respective subproblems can be solved
efficiently. We can ensure this in both cases.

For stochastic load balancing on unrelated machines, each job creates at most~$m$ configurations which trivially implies that the subproblems can be solved efficiently. 
Here, the LP relaxation of configuration balancing used in \Cref{thm_config_off} is equivalent to the LP relaxation of the generalized assignment problem (GAP) 
solved in~\cite{DBLP:journals/mp/ShmoysT93}. Hence,~\Cref{thm_config_off} implies the following theorem. 

\begin{theorem}\label{thm_unrel_lb_off}
	There exist  efficient deterministic algorithms that compute a non-adaptive policy for load balancing on unrelated machines with stochastic jobs that achieve an $\Theta\big(\frac{\log m}{\log \log m}\big)$-approximation \emph{offline} and an $\Theta(\log m)$-approximation \emph{online} when comparing to the optimal adaptive policy.
\end{theorem}

These results are asymptotically tight as shown by the lower bound of~$\Omega\big(\frac{\log m}{\log \log m}\big)$ on the adaptivity gap~\cite{DBLP:journals/mor/Gupta0NS21} and the lower bound of~$\Omega(\log m)$ on the competitive ratio of any deterministic online algorithm, even for deterministic requests~\cite{DBLP:journals/jal/AzarNR95}. In particular, the theorem implies that the adaptivity gap for stochastic load balancing is $\Theta\big(\frac{\log m}{\log \log m}\big)$.

For virtual circuit routing, efficiently solving the subproblems requires more work as the configurations are only given \emph{implicitly}. For the offline setting, since the LP relaxation has (possibly) exponentially many variables, we design an efficient separation oracle for the dual LP in order to efficiently solve the primal. For the online setting, we carefully select a subset of polynomially many configurations that contain the configuration chosen by the greedy algorithm, even when presented with all configurations. Thus, \Cref{thm_config_off} implies that stochastic requests are not harder to approximate than deterministic requests in both settings.

\begin{theorem}\label{thm_route_off}\label{thm_routing}
    For routing with stochastic requests when comparing to  the optimal adaptive policy there exists an efficient randomized \emph{offline} algorithm that computes an  $\Theta\big(\frac{\log m}{\log \log m}\big)$-approximation and there exists  an efficient  \emph{online} algorithm that computes an $\Theta(\log m)$-approximation.
\end{theorem}

\paragraph{Adaptive policies for related machines.}

In the case of related machines, we improve on the above result by using adaptivity. 
\begin{theorem}\label{thm_rel_lb_off}\label{thm_related}
For load balancing on related machines with stochastic jobs when comparing to  the optimal  adaptive policy there exist an efficient \emph{offline} algorithm that computes an adaptive 
$O(1)$-approximation and an efficient \emph{online} algorithm that computes an $O(\log \log m)$-approximation.
\end{theorem}

It remains an open question whether the online setting admits an $O(1)$-competitive algorithm.

As adaptivity turns out to be very powerful in load balancing on related machines, it is reasonable to ask whether the (stochastic) information about job sizes is even needed to improve upon \Cref{thm_config_off}. In~\Cref{appendix_nonclairvoant_related}, we answer this question to the affirmative in the following sense. We show that {\em non-clairvoyant algorithms}, that have no prior knowledge of the job sizes, approximate the optimal offline schedule only within a factor $\Omega(\sqrt{m})$, even if the size of a job is revealed immediately upon assigning it to a machine.
Further, notice that \Cref{thm_config_off} shows that even non-adaptive policies can breach this non-clairvoyance barrier when given stochastic job sizes. 

\subsection{Technical Overview}

We illustrate the main idea behind our non-adaptive policies,
  which compare to the optimal offline adaptive policy.
As in many other stochastic optimization problems, our goal is to give a good deterministic proxy for the makespan of a policy. Then, our algorithm will optimize over this deterministic proxy to obtain a good solution. First, we observe that if all configurations were bounded with respect to $\mathbb{E}[\Opt]$ in every entry, then selecting configurations such that each resource has expected load $O(\mathbb{E}[\Opt])$ gives the desired $O\big(\frac{\log m}{\log \log m}\big)$-approximation by standard concentration inequalities for independent sums with bounded increments. Thus, in this case the expected load on each resource is a good proxy.
However, in general, we have no upper bound on $X_{ij}(c)$, so we cannot argue as above. We turn these unbounded random variables into bounded ones in a standard way by splitting each request into \emph{truncated} and \emph{exceptional} parts. 

\begin{definition}[Truncated and Exceptional Parts]\label{def_truncated_and_exceptional}
	Let $\tau \geq 0$ be a fixed threshold. For a random variable~$X$, its truncated part (with respect to threshold~$\tau$) is~$X^T := X \cdot \one_{X < \tau}$. Similarly, its exceptional part is~$X^E := X \cdot \one_{X \geq \tau}$. Note that~$X = X^T + X^E$. 
\end{definition}

It is immediate that the truncated parts $X_{ij}^T(c)$ are bounded in $[0, \tau]$. Taking $\tau = O(\EE[\Opt])$, we can control their contribution 
to the makespan using concentration. It remains to find a good proxy for the contribution of exceptional parts to the makespan. This is one of the main technical challenges of our work as we aim to compare against the optimal adaptive policy. We will see that adaptive policies have much better control over the exceptional parts than non-adaptive ones.

Concretely, let~$c_j$ be the configuration chosen by some fixed policy for request~$j$. Note that~$c_j$ itself can be a random variable in~$\{1,\ldots,q_j\}$. We want to control the quantity 
\[
\mathbb{E}\Big[\max_i \sum_{j=1}^n  X_{ij}^E(c_j) \Big].
\]
Because we have no reasonable bound on the $X_{ij}^E(c_j)$'s, for non-adaptive policies, we can only upper bound the expected maximum by the following sum
\begin{align}\label{eq_exceptional_proxy}
	\EE\Big[\max_{1 \leq 1 \leq m} \sum_{j=1}^n  X_{ij}^E(c_j) \Big] 
	\leq \sum_{j=1}^n  \EE \Big[\max_{1 \leq i \leq m} X_{ij}^E(c_j) \Big].
\end{align}
We call the right hand side 
\emph{total (expected) exceptional load}. The above inequality is tight up to constants for non-adaptive policies, so it seems like the total expected exceptional load is a good proxy to use for our algorithm. However, it is far from tight for adaptive policies as 
the~example~shows.

\begin{example}\label{ex_except_lower}
	We define an instance of load balancing on \emph{related} machines, 
 where each machine~$i$ has a speed $s_i$ and each job $j$ has processing time $X_j$ such that $X_{ij} = \frac{X_j}{s_i}$.
	Our instance has one ``fast'' machine with speed $1$ and $m-1$ ``slow'' machines each with speed $\frac{1}{\tau m}$, where $\tau > 0$ is the truncation threshold. There are $m$ jobs: a stochastic one with processing time~$\tau \cdot$Ber$\big(\frac{1}{\tau}\big)$ and $m-1$ deterministic jobs with processing time $\frac{1}{m}$.
	The optimal adaptive policy schedules first the stochastic job on the fast machine. If its realized size is $0$, then it schedules all deterministic jobs on the fast machine. Otherwise its realized size is $\tau$ and we schedule one deterministic job on each slow machine. This gives $\mathbb{E}[\Opt] = \big(1 - \frac{1}{\tau}\big)\big(\frac{m-1}{m}\big) + \frac{1}{\tau} \cdot \tau = \Theta(1)$. However, the total expected exceptional load (with respect to threshold~$\tau$) is $\sum_{i,j} \mathbb{E} \big[X_{ij}^E \cdot \one_{j \rightarrow i}\big] = \frac{1}{\tau}(m \tau) = m$.
\end{example}

In the example, the optimal adaptive policy accrues a lot of exceptional load, but this does not have a large affect on the makespan. Concretely, \eqref{eq_exceptional_proxy} can be loose by a $\Omega(m)$-factor for adaptive policies. Thus, it seems that the total exceptional load is a bad proxy in terms of lower-bounding $\Opt$. However, we show that, by comparing our algorithm to a \emph{near-optimal} adaptive policy rather than the optimal one, the total exceptional load becomes a good proxy in the following sense. This is the main technical contribution of our work, and it underlies all of our algorithmic techniques.

\begin{restatable}{theorem}{thmadaprestart}\label{thm_adap_restart}
    For configuration balancing with stochastic requests, there exists an adaptive policy with expected maximum load and total expected exceptional load at most $2 \cdot \mathbb{E} [\Opt]$ with respect to any truncation threshold $\tau \geq 2 \cdot \mathbb{E}[\Opt]$. Further, any configuration~$c$ selected by this policy satisfies~$\EE\big[\max_i X_i(c)\big] \leq \tau$. 
\end{restatable}

The proof of the above relies on carefully modifying the ``decision tree'' representing the optimal adaptive policy. In light of \Cref{thm_adap_restart}, the deterministic proxies we consider are the expected truncated load on each resource and the total expected exceptional load. All of our algorithms then proceed by ensuring that both quantities are bounded with respect to $\EE[\Opt]$. In the offline case, we round a natural assignment-type linear program (LP), and in the online case, we use a potential function argument. All of these algorithms actually output non-adaptive policies.

For the special case of related-machines load balancing, we also compute a non-adaptive assignment but instead of following it exactly, we deviate using adaptivity and give improved solutions.


\subsection{Related Work}

While stochastic optimization problems have long been
studied~\cite{beale55,dantzig55}, approximation algorithms for them
are more recent~\cite{DBLP:journals/jacm/MohringSU99,DyeST2003}. By
now, multi-stage stochastic problems (where uncertain information is
revealed in stages) are
well-understood~\cite{CharikarCP2005,GuptaPRS2011,SwamyS2012}. In
contrast, more dynamic models, where the exact value of an unknown
parameter becomes known at times depending on the algorithms decisions
(serving a request) still remain poorly
understood. 
Some exceptions come from stochastic
knapsack~\cite{BhalgatGK11,deanGV08,GuptaKMR11,Ma18} as well as
stochastic scheduling and routing which we discuss below.

\medskip \noindent {\bf Scheduling.\ \ } For load
balancing with deterministic sizes, a $2$-approximation in the most general
unrelated-machines offline
setting~\cite{DBLP:journals/mp/LenstraST90} is known. For identical machines
($p_{ij} = p_j$ for all jobs~$j$), the greedy algorithm (called \emph{list scheduling}) is a $\big(2 - \frac1m\big)$-approximation
algorithm~\cite{DBLP:journals/siamam/Graham69}. This guarantee holds
even when the jobs arrive online and \emph{nothing} is known about job
sizes. This implies a~$\big(2 - \frac1m\big)$-approximate
\emph{adaptive} policy for stochastic load balancing on identical
machines.

Apart from this, prior work on stochastic scheduling has focused on
approximating the optimal \emph{non-adaptive} policy. There are non-adaptive $O(1)$-approximations known for identical machines~\cite{DBLP:journals/siamcomp/KleinbergRT00}, unrelated machines~\cite{DBLP:journals/mor/Gupta0NS21} and the $\ell_q$-norm objective~\cite{Molinaro19}.
\cite{DBLP:journals/mp/GuptaKNS22} give  non-adaptive $\poly(\log \log m)$-approximations for more general load balancing problems, where at least~$t$ stochastic sets out of a structured set system of size $n$ have to be selected. 
  
In contrast, our work  focuses on approximating the stronger optimal
\emph{adaptive} policy. The \emph{adaptivity gap} (the ratio between
the expected makespan of the optimal adaptive and non-adaptive
policies) can be $\Omega\big(\frac{\log m}{\log \log m}\big)$ even for the simplest case of identical machines \cite{DBLP:journals/mor/Gupta0NS21}. Thus, previous work on approximating the optimal non-adaptive policy does not immediately give any non-trivial approximation guarantees for our setting.  The only previous work on adaptive stochastic policies for
load-balancing (beyond the highly-adaptive list scheduling) is by~\cite{SagnolW21}. They propose scheduling policies whose degree of adaptivity can be controlled by parameters and show an approximation factor of $O(\log \log m)$ for scheduling on identical machines. 

Other objectives have been studied in stochastic scheduling, namely minimizing the expected sum of (weighted) completion times. 
Most known adaptive 
policies have an approximation ratio depending on parameters of the distribution~\cite{GuptaMUX20,MegowUV06,MohringSU99,schulz08,SkutellaU01,SkutellaSU16}, with the notable (still polylogarithmic) exception in~\cite{ImMP15}. 

Online load balancing with deterministic jobs is also well studied~\cite{Azar96}. On identical machines, the aforementioned list scheduling algorithm~\cite{DBLP:journals/siamam/Graham69} is~$\big(2 - \frac1m\big)$-competitive. For unrelated machines, there is a deterministic $O(\log m)$-competitive algorithm \cite{AspnesAFPW97} and this is best possible~\cite{DBLP:journals/jal/AzarNR95}. When the machines are uniformly related, \cite{BermanCK00} design an $O(1)$-competitive algorithm for minimizing the makespan. \cite{ImKPS18} and \cite{ImKKP19} study the multi-dimensional generalization to vector scheduling under the makespan and the~$\ell_q$-norm objective.

To the best of our knowledge, configuration balancing has not been
  explicitly defined before. 
  The techniques
  of~\cite{AspnesAFPW97} give an $O(\log m)$-competitive algorithm
  for deterministic requests; it is also studied for packing integer
  programs in the random order setting~\cite{AD15,AWY14,GM-MOR16}.

\medskip \noindent {\bf Routing.\ \ } For stochastic routing 
there is hardly anything known, except rather specialized results for a packing version~\cite{ChawlaR06,GuptaK17}. Here, the task is to (adaptively) select  a subset of uncertain demands to be routed between pairs of nodes in a given graph such that the total demand on an edge does not exceed a given edge capacity and the total value of the successfully routed demand is maximized in expectation. 
Adaptive policies with logarithmic approximation ratios are known for routing requests with a single sink in planar~\cite{ChawlaR06} and arbitrary directed graphs~\cite{GuptaK17}. Further, there is a constant non-adaptive policy for routing in trees~\cite{GuptaK17}. The packing flavor of this problem is very different from configuration balancing where {\em all} requests must be served.

The offline variant of virtual circuit routing is mostly referred to by congestion minimization. When~$d_j = 1$ for each source-sink pair~$(s_j,t_j)$, there is an $O\big( \frac{\log m}{\log \log m}\big)$-approximation algorithm by~\cite{RaghavanT87}, 
which has been shown to be best possible, unless~$\textup{NP} \subseteq \textup{ZPTIME}(n^{\log \log n})$~\cite{ChuzhoyGKT07}. 

In the online setting, when the source-sink pairs arrive online over a
list and have to be routed before the next pair
arrives,~\cite{AspnesAFPW97} give a lower bound of~$\Omega(\log n)$ on
the competitive ratio of any deterministic online algorithm in
directed graphs, where~$n$ is the number of vertices. They also give a
matching upper bound. 
For more details on online routing we refer to the survey 
\cite{Leonardi96}. 



\section{Configuration Balancing with Stochastic Requests}\label{sec_config_balancing}

In this section, we prove our main results for the most general problem we consider: configuration balancing. We give a $O\big(\frac{\log m}{\log \log m}\big)$-approximation offline and a $O(\log m)$-approximation online. Both of our algorithms are non-adaptive. Before describing the algorithms, we give our main structure theorem that enables all of our results. Roughly, we show that instead of comparing to the optimal adaptive policy, by losing only a constant factor in the approximation ratio, we can compare to a near-optimal policy that behaves like a non-adaptive one (with respect to the proxy objectives we consider -- namely, the total expected exceptional load).

\subsection{Structure theorem: ``Near optimal policies are almost non-adative''}

The goal of this section is to show that there exists a near-optimal policy as guaranteed by \Cref{thm_adap_restart}.
To this end, we modify the optimal policy by ``restarting'' whenever an exceptional request is encountered. Additionally, we ensure that this modified policy never selects a configuration~$c$ for a request~$j$ with~$\EE\big[\max_i X_{ij}(c) \big] > \tau$. 

We let $J$ denote the set of requests. For any subset $J' \subseteq J$, we let $\Opt(J')$ denote the optimal adaptive policy (and its maximum load) on the set of requests $J'$. Note that~$\Opt(\emptyset) = 0$. Our (existential) algorithm to construct such a policy will begin by running the optimal policy $\Opt(J)$ on all requests. However, once a exceptional request is encountered or the next decision will choose a configuration with too large expected max, we cancel $\Opt(J)$ and instead recurse on all remaining requests, ignoring all previously-accrued loads. They main idea of our analysis is that we recurse with small probability. We now proceed formally.

\thmadaprestart* 

\begin{proof}
    We prove the theorem by induction on the number of requests $n \in \mathbb N$. The base case $n = 0$ is trivial. Now we consider~$n>0$. Let $J$ be the set of $n$ requests. Our algorithm to construct the desired policy $S(J)$ is the following. Throughout, we fix a truncation threshold $\tau \geq 2\cdot \EE[\Opt]$.

\begin{algorithm}
    \DontPrintSemicolon
	\caption{Policy~$S(J)$}
	\label{alg_near_opt_policy}
	 \While{$J \neq \emptyset$}{
	    $c_j \leftarrow$ configuration chosen for next request~$j$ by \Opt(J) \tcp*[r]{run $\Opt(J)$}
	    \nl\label{alg_L} \uIf(\tcp*[f]{maximum too large}){$\EE\big[\max_i X_{ij}(c_j) \big] > \tau$}{
	        {stop} \;
	    }
	    \Else{
	      choose~$c_j$ for request~$j$ \;
	      $J \leftarrow J \setminus \{ j\}$ \; 
	      \nl\label{alg_E} \If(\tcp*[f]{exceptional configuration}){$\max_{i} X_{ij}(c_j) \geq \tau$}{
	        {stop} \;
	      }
	    }
	}
	run $S(J)$ \tcp*[r]{recurse with remaining requests}
\end{algorithm}

	Let $R$ be the random set of requests we recurse on after stopping $\Opt(J)$. We first show that indeed $\lvert R \rvert < \lvert J \rvert$, so we can apply induction. Suppose we did not follow $\Opt(J)$ to completion because a chosen configuration becomes exceptional (\ref{alg_E}); denote this event by~$\mathcal E$. In this case, there is at least one request for which we have chosen a configuration. Hence, we have~$|R| < |J|$, and therefore there is a policy~$S$ with the required properties by induction. 
    
    Suppose now that~$\Opt(J)$ chooses a configuration~$c_j$ for request~$j$ that is too large (\ref{alg_L}); denote this event by~$\mathcal L$. We have to show that~$|R| < |J|$ holds as well. Suppose for the sake of contradiction that~$j$ was the first request considered by~$\Opt(J)$. As \Opt{} is w.l.o.g. deterministic, this implies~$\EE[\Opt] \geq \EE\big[\max_i X_{ij}(c_j) \big] > 2 \EE[\Opt]$, a contradiction. Hence, the desired policy~$S$ exists by induction. 

    The maximum load of this policy is at most $\Opt(J) + S(R)$, where we set $R = \emptyset$ if no decision results in an exceptional or too large configuration when running $\Opt(J)$. 
    In expectation, we have
        \begin{align*}
            \mathbb{E} [S(R)] & = 
            \sum_{J' \subsetneq J} \mathbb{E}[S(R) \mid R = J'] \mathbb{P}[R = J'] = \sum_{ J' \subsetneq J} \mathbb{E} [S(J')] \mathbb{P}[R = J'] \leq 2 \sum_{J' \subsetneq J} \mathbb{E} [\Opt(J')] \mathbb{P}[R = J] \\
            & \leq 2 \cdot \mathbb{E} [\Opt(J)] \mathbb{P}[R \neq \emptyset].
        \end{align*}
    In the second equality, we use the fact that the realizations of the remaining requests in $R$ are independent of the event $R = J'$. The first inequality uses the inductive hypothesis. The last inequality uses $J' \subseteq J$, so $\mathbb{E} [\Opt(J')] \leq \mathbb{E} [\Opt(J)]$, and $\Opt(\emptyset) = 0$. 
    
    Note that on the event $R \neq \emptyset$, we have that $\Opt(J)$ chooses a configuration that becomes exceptional or that is too large in expectation. By definition of the policy~$S$, the events~$\mathcal E$ and~$\mathcal L$ are disjoint. 
    By definition of~$\mathcal E$, we have~$\Opt(J)\cdot \one_{\mathcal E}  \geq \tau \cdot \one_{\mathcal E} $. 
    Observe that the event~$\mathcal L$ implies that there is a request~$j^*$ with configuration~$c^*$ chosen by~$\Opt(J)$ with~$\EE\big[\max_i X_{ij^*}(c^*) \big] \geq \tau$. Since the realization of~$\max_i X_{ij^*} (c^*)$ is independent of the choice~$c^*$, this implies~$\EE[\Opt \mid \mathcal L ] \geq \EE[\max_i X_{ij^*}(c^*) \mid \mathcal L] = \EE[\max_i X_{ij^*}(c^*)] \geq \tau$. Thus, 
    \begin{equation*}
        \EE[\Opt(J)] \geq \mathbb P[ \mathcal E] \EE[\Opt(J) \mid \mathcal E ] + \mathbb P [ \mathcal L]  \EE[\Opt(J) \mid \mathcal L ] \geq \mathbb P[ \mathcal E] \tau  + \mathbb P [ \mathcal L] \tau = 2 \mathbb P[R \neq \emptyset ] \EE[\Opt(J)].
    \end{equation*}
    Rearranging yields $\mathbb{P}[R \neq \emptyset] \leq \frac{1}{2}$. 
    Hence, we can bound the expected makespan of policy~$S(J)$ by
        \[\mathbb{E} [\Opt(J)] + \mathbb{E} [S(R)] \leq \mathbb{E} [\Opt(J)] + 2  \mathbb{E} [\Opt(J)]  \mathbb{P}[R \neq \emptyset] \leq 2 \mathbb{E} [\Opt(J)].\]
    
    
        
    The computation for the total expected exceptional load is similar. We let $j \rightarrow c$ denote the event that our policy chooses configuration $c$ for request $j$. Then, we can split the exceptional load into two parts based on whether a configuration is chosen by $\Opt(J)$ or $S(R)$
    \[\sum_{j=1}^n \sum_{c=1}^{q_j} \Big(\max_i X_{ij}^E(c) \Big) \cdot \one_{j \rightarrow c} = \sum_{j=1}^n \sum_{c=1}^{q_j}\Big(\max_i X_{ij}^E(c) \Big) \cdot \one_{j \xrightarrow{J} c} + \sum_{j=1}^n \sum_{c=1}^{q_j} \Big(\max_i X_{ij}^E(c) \Big) \cdot \one_{j \xrightarrow{R} c},\]
    where we let $j \xrightarrow{J} c$ and $j \xrightarrow{R} c$ denote the events that configuration $c$ is chosen for request $j$ by $\Opt(J)$ up to the first too large configuration or up to and including the first exceptional configuration, or by $S(R)$, respectively.
        
    We first bound the former term, corresponding to the configurations chosen in $\Opt(J)$. In case of event~$\mathcal L$ or if~$\Opt(J)$ is run to completion, we have $\sum_{j,c} \big(\max_i X_{ij}^E(c) \big) \cdot \one_{j \xrightarrow{J} c} = 0$. 
    Otherwise, let $j^* \rightarrow c^*$ be the first (and only) exceptional configuration chosen by $\Opt(J)$. Then, $\sum_{j,c} \big(\max_i X_{ij}^E(c) \big) \cdot \one_{j \xrightarrow{J} c} = \max_i X_{ij^*}^E(c^*) \leq \Opt(J)$. Combining and taking expectations yields
    \[
        \mathbb{E} \bigg[\sum_{j=1}^n \sum_{c=1}^{q_j} \max_i X_{ij}^E(c) \cdot \one_{j \xrightarrow{J} c}\bigg] \leq \mathbb{E} [\Opt(J)].
    \]
        
    For the latter term, we condition again on the events $R = J'$ and apply the inductive hypothesis. All exceptional parts are defined with respect to the fixed threshold $\tau \geq 2 \cdot \mathbb{E} [\Opt(J)] \geq 2 \cdot \mathbb{E} [\Opt(J')]$ for $J' \subset J$. Therefore, 
    \begin{align*}
        \mathbb{E} \bigg[\sum_{j=1}^n \sum_{c=1}^{q_j} \big(\max_i X_{ij}^E(c) \big) \cdot \one_{j \xrightarrow{R} c}\bigg] &= \sum_{ J' \subsetneq J} \mathbb{E} \bigg[\sum_{j \in J'} \sum_{c=1}^{q_j} \big(\max_i X_{ij}^E(c) \big) \cdot \one_{j \xrightarrow{R} c} \,\Big|\, R = J'\bigg] \cdot \mathbb{P}[R = J']\\
        &= \sum_{ J' \subsetneq J} \mathbb{E}\bigg[ \sum_{j \in J'} \sum_{c=1}^{q_j} \big(\max_i X_{ij}^E(c) \big) \cdot \one_{j \xrightarrow{J'} c}\bigg] \cdot \mathbb{P}[R = J']\\
        &\leq 2 \sum_{ J' \subsetneq J} \mathbb{E} [\Opt(J')] \cdot \mathbb{P}[R = J']\\
        &\leq 2 \cdot \mathbb{E} [\Opt(J)] \cdot \mathbb{P}[R \neq \emptyset] \leq \mathbb{E} [\Opt(J)].
    \end{align*}
        Note that we define $j \xrightarrow{J'} c$ to be the event that the policy $S(J')$ chooses configuration~$c$ for request~$j$. In conclusion, by combining our bounds for these two terms we have
        \begin{align*}
            \mathbb{E} \bigg[\sum_{j=1}^n \sum_{c=1}^{q_j} \big(\max_i X_{ij}(c)^E \big) \cdot \one_{j \rightarrow c} \bigg] 
            & \leq 2 \mathbb{E} [\Opt(J).
        \end{align*}
        To conclude, our constructed policy has expected makespan and total expected exceptional load both at most $2 \cdot \EE [\Opt{}]$ (by the above calculations), and it never chooses a configuration with $\EE \big[ \max_i X_i(c) \big] > \tau$ (because we stop running $\Opt(J)$ right before it chooses such a configuration, and by induction we subsequently do not as well.)
\end{proof}

Having this near-optimal policy at hand, the upshot is that we can bound our subsequent algorithms with respect to the following LP relaxation \eqref{eq_lpstochconfig} for configuration balancing with stochastic requests. The variable~$y_{cj}$ denotes selecting configuration~$c$ for request~$j$. We take our threshold between the truncated and exceptional parts to be~$\tau$. Using the natural setting of the $y$-variables defined by the policy guaranteed by \Cref{thm_adap_restart}, it is straight-forward to show that the following LP relaxation is feasible, formalized in \Cref{lem_lpc_feasible} and proven in \Cref{app_config}. 
%
\begin{equation}\tag{$\lpconfig$}\label{eq_lpstochconfig}
	\begin{array}{rll}
		\sum_{c=1}^{q_j} y_{cj} &  = 1 &\quad\forall~j\in[n] \\
		\sum_{j=1}^n \sum_{c=1}^{q_j} \EE[X_{ij}^T(c)] \cdot y_{cj}  &  \leq \tau &\quad\forall~i\in[m] \\
		\sum_{j=1}^n \sum_{c=1}^{q_j} \EE[\max_i X_{ij}^E(c)] \cdot y_{cj} & \leq \tau \\
		y_{cj} & = 0 & \quad \forall~ j \in [n], \forall~c \in [q_j]: \EE[\max_i X_{ij}(c)] > \tau \\ 
		y_{cj} & \geq 0 & \quad \forall~j\in[n], \forall~c \in [q_j] 
	\end{array} 
\end{equation}
%
\begin{restatable}{lem}{lemlpfeasible}\label{lem_lpc_feasible}
    \eqref{eq_lpstochconfig} is feasible for any $\tau \geq 2 \cdot \mathbb{E}[\Opt]$.
\end{restatable}

\subsection{Offline Setting}

Our offline algorithm is based on the natural randomized rounding of \eqref{eq_lpstochconfig}. For the truncated parts, we use the following maximal inequality to bound their contribution to the makespan. See \Cref{app_prelim} for proof. The independence is only required for the random variables constituting a particular sum~$S_i$, but is not necessary for random variables appearing in different sums.

\begin{restatable}{lem}{lemmaxbound}\label{lem_max_bound}
	Let $S_1, \dots, S_m$ be sums of independent random variables, that are bounded in $[0,\tau]$ for some $\tau > 0$, such that $\mathbb{E} [S_i] \leq \tau$ for all $1 \leq i \leq m$. Then, $\mathbb{E} [\max_i S_i] = O\big(\frac{\log m}{\log \log m}\big)\tau$.
\end{restatable}

To bound the contribution of the exceptional parts, we use \eqref{eq_exceptional_proxy} (i.e. the total expected exceptional load.)
%
Using binary search for the correct choice of~$\tau$ and re-scaling the instance by the current value of~$\tau$, it suffices to give an efficient algorithm that either
\begin{itemize}[nolistsep]
	\item outputs a non-adaptive policy with expected makespan $O\big(\frac{\log m}{\log \log m}\big)$, or
	\item certifies that $\mathbb{E} [\Opt] > 1$.
\end{itemize}


To that end, we use natural independent randomized rounding of \eqref{eq_lpstochconfig}. That is, if \eqref{eq_lpstochconfig} has a feasible solution~$y^*$, for request~$j$, we choose configuration~$c$ as configuration~$c_j$ independently with probability~$y_{cj}^*$; see 
\Cref{alg_off_config_stoch}.
\begin{algorithm}
	\DontPrintSemicolon
	\caption{Offline Configuration Balancing with Stochastic Requests}
	\label{alg_off_config_stoch}
	try to solve  \eqref{eq_lpstochconfig} with $\tau = 2$\;
	\uIf{\eqref{eq_lpstochconfig} is feasible}{
		let $y^*$ be the outputted feasible solution\;
		\For{each request $j$}{
			independently sample $c \in [q_j]$ with probability $y^*_{cj}$ \;
			choose sampled $c$ as $c_j$ \;
		}
	}
	\Else{
		return ``$\mathbb{E} [\Opt] > 1$''\;
	}
\end{algorithm}
If the configurations are given explicitly as part of the input, then  \eqref{eq_lpstochconfig} can be solved in polynomial time and, thus, \Cref{alg_off_config_stoch} runs in polynomial time. Hence, the desired $O\big(\frac{\log m}{\log \log m}\big)$-approximate non-adaptive policy for configuration balancing with stochastic requests (\Cref{thm_config_off}) follows from the next lemma.

\begin{lem}\label{lem_off_config_stoch}
	If \eqref{eq_lpstochconfig} can be solved in polynomial time, \Cref{alg_off_config_stoch} is a polynomial-time randomized algorithm that either outputs a non-adaptive policy with expected makespan $O\big(\frac{\log m}{\log \log m}\big)$, or certifies correctly that $\mathbb{E} [\Opt] > 1$.
\end{lem}
\begin{proof}
	We need to show that \Cref{alg_off_config_stoch} either outputs a non-adaptive policy with expected makespan $O\big(\frac{\log m}{\log \log m}\big)$ or certifies that $\mathbb{E} [\Opt] > 1$. There are two cases.
	
	If \eqref{eq_lpstochconfig} is feasible for $\tau = 2$, then we output a (randomized) non-adaptive policy. We show this policy has expected makespan $O\big(\frac{\log m}{\log \log m}\big)$. We let $j \rightarrow c$ denote the event that our algorithm chooses configuration~$c$ for request~$j$. Thus, the truncated load on resource~$i$ can be written as
	\[L_i = \sum_{j=1}^{n} \bigg( \sum_{c=1}^{q_j} X^T_{ij}(c) \cdot \one_{j \rightarrow c} \bigg).\]
	Note that the random variables $\sum_{c=1}^{q_j} X_{ij}^T(c) \cdot \one_{j \rightarrow c}$ are independent for different~$j$ because the~$X_{ij}$ are and we sample the configuration~$c_j$ independently for each~$j$. Further, they are bounded in $[0,2]$ by truncation. With the constraints of \eqref{eq_lpstochconfig}, we can bound the expectation by
	\[ \EE [L_i] = \sum_{j=1}^{n} \sum_{c=1}^{q_j} \EE\Big[X_{ij}^T(c) \cdot \one_{j \rightarrow c}\Big] = \sum_{j=1}^{n} \sum_{c=1}^{q_j} \EE\Big[X_{ij}^T(c)\Big] \cdot y^*_{cj} \leq 2,\]
	where we used the independence of~$\one_{j \rightarrow c}$ and~$X_{ij}(c)$ in the second equality. 
	By \Cref{lem_max_bound}, $\EE [\max_i L_i] = O\big(\frac{ \log m}{\log \log m}\big)$. Using \eqref{eq_exceptional_proxy} we upper bound the total expected exceptional load by
	\[\EE \bigg[\max_{1\leq i \leq m} \sum_{j=1}^{n} \sum_{c=1}^{q_j} X^E_{ij}(c) \cdot \one_{j \rightarrow c} \bigg] \leq \sum_{j=1}^{n} \sum_{c=1}^{q_j} \EE \Big[ \max_{1\leq i \leq m} X^E_{ij}(c) \cdot \one_{j \rightarrow c} \Big] = \sum_{j=1}^{n} \sum_{c=1}^{q_j} \EE\Big[\max_{1\leq i \leq m} X^E_{ij}(c)\Big] \cdot y_{cj}^* \leq 2.\]
	Combining our bounds for the truncated and exceptional parts completes the proof. The expected makespan of our algorithm is given by
	\[\EE \Big[\max_{1\leq i \leq m} \sum_{j=1}^{n} \sum_{c=1}^{q_j} X_{ij}(c) \cdot \one_{j \rightarrow c} \Big] \leq \EE \Big[\max_{1\leq i \leq m} L_i \Big] + \EE \Big[\max_{1\leq i \leq m} \sum_{j=1}^{n} \sum_{c=1}^{q_j} X^E_{ij}(c) \cdot \one_{j \rightarrow c} \Big] = O\Big(\frac{ \log m}{\log \log m}\Big).\]
	
	In the other case \eqref{eq_lpstochconfig} is infeasible, so the contrapositive of \Cref{lem_lpc_feasible} gives $\EE [\Opt] > 1$.
\end{proof}

\subsection{Online Setting} 

We now consider online configuration balancing where~$n$ stochastic requests arrive online one-by-one, and for each request, one configuration has to be irrevocably selected before the next request appears. We present a non-adaptive online algorithm 
that achieves a competitive ratio of~$O(\log m)$, which is best possible due to the lower bound of~$\Omega(\log m)$ on the competitive ratio of any deterministic algorithm for online load balancing on unrelated machines~\cite{DBLP:journals/jal/AzarNR95}. 

First, by a standard guess-and-double scheme, we may assume we have a good guess of $\EE[\Opt]$.

\begin{lem}\label{lem_guess_and_double}
	Given an instance of online configuration balancing with stochastic requests, suppose there exists an online algorithm that, given parameter $\lambda > 0$, never creates an expected makespan more than $\alpha \cdot \lambda$, possibly terminating before handling all requests. Further, if the algorithm terminates prematurely, then it certifies that $\EE [\Opt] > \lambda$.  Then, there exists an $O(\alpha)$-competitive algorithm for online configuration balancing with stochastic requests. Further, the resulting algorithm preserves non-adaptivity.
\end{lem}
\noindent We omit the proof, which is analogous to its virtual-circuit-routing counterpart in \cite{AspnesAFPW97}.


We will build on the same technical tools as in the offline case. In particular, we wish to compute a non-adaptive assignment online with small expected truncated load on each resource and small total expected exceptional load. To achieve this, we generalize the greedy potential function approach of \cite{AspnesAFPW97}. Our two new ingredients 
are to treat the exceptional parts of a request's configuration as a resource requirement for an additional, artificial resource and to compare the potential of our solution directly with a \emph{fractional} solution to \eqref{eq_lpstochconfig}. 

Now we describe our potential function, which is based on an exponential/soft-max function. Let~$\lambda$ denote the current guess of the optimum as required by \Cref{lem_guess_and_double}. We take~$\tau = 2 \lambda$ as our truncation threshold. Given load vector~$L \in \RR^{m+1}$, our potential function is 
\[
\phi(\load) = \sum_{i=0}^m (3/2)^{ \load_i / \tau }.
\] 
For $i = 1, \dots, m$, we will ensure the $i$th entry of $\load$ is the \emph{expected} truncated load on resource~$i$. We use the $0$th entry as a virtual resource that is the total expected exceptional load. For any request~$j$, we let $\load_{ij}$ be the $i$th entry of the expected load vector after handling the first $j$ requests. We define $\load_{i0} := 0$ for all $i$. Let $L_j$ be the expected load vector after handling the first $j$ requests.

\Cref{alg_on_stoch_config} works as follows: Upon arrival of request~$j$, we try to choose the configuration~$c_{j} \in [q_{j}]$ that minimizes the increase in potential. Concretely, we choose $c_j$ to minimize \[
\Big((3/2)^{( \load_{0j-1} + \EE[\max_{i \in [m]} X_{ij}^E(c_{j})] ) / \tau    } + 
\sum_{i=1}^m (3/2)^{ ( \load_{ij-1} + \EE[X_{ij}^T(c_{j})])    / \tau }\Big)         - \phi(\load_{j-1}).
\]
%
%
\begin{algorithm}
	\DontPrintSemicolon
	\caption{Online Configuration Balancing with Stochastic Requests}
	\label{alg_on_stoch_config}  
	$\ell \leftarrow \log_{3/2} (2m + 2)$ \;
	$\lambda \leftarrow $ current guess of~$\EE[\Opt]$ \;
	$\tau \leftarrow 2 \lambda $ truncation threshold \;
    \Upon{arrival of request $j$}{
    	\nl\label{alg_on_config_argmin} $c_{j} \leftarrow \arg \min_{c \in [q_j]} \Big((\nicefrac32)^{( \load_{0j-1} + \EE[\max_{i \in [m]} X_{ij}^E(c)]) / \tau } + 
    	\sum_{i=1}^m (\nicefrac32)^{ ( \load_{ij-1} + \EE[X_{ij}^T(c)]) / \tau }\Big)         - \phi(\load_{j-1})$\;
    	\uIf{
    		$\load_{ij-1} + \EE[X_{ij}^T(c_{j})] \leq \ell \tau $  for all $i \in [m]$ {\normalfont \bfseries and }$\load_{0j-1} + \EE[\max_{i \in [m]} X_{ij}^E(c_j)] \leq \ell \tau $ 
    	}{
    		choose $c_j$ for $j$ \;
    		$\load_{ij} \leftarrow \load_{ij-1} + \EE[X_{ij}^T(c_{j})]$ for all $i \in [m]$\; 
    		$\load_{0j} \leftarrow \load_{0j-1} + \EE[\max_{i \in [m]} X_{ij}^E(c_j)]$ \;     }
    	\Else{
    		return ``$\mathbb{E} [\Opt] > \lambda$'' \;
    	}
	}
\end{algorithm}


To analyze this algorithm, we compare its makespan with a solution to~\eqref{eq_lpstochconfig}. 
This LP has an integrality gap of~$\Omega\big(\frac{\log m}{\log \log m}\big)$, which follows immediately from the path assignment LP for virtual circuit routing \cite{LeightonRS98}. Hence, a straightforward analysis of \Cref{alg_on_stoch_config} comparing to a rounded solution to \eqref{eq_lpstochconfig} gives an assignment with expected truncated load per machine and total expected exceptional load $O\big(\log m \cdot \frac{\log m}{\log \log m}) \cdot \EE[\Opt]$. The $O\big(\frac{\log m}{\log \log m}\big)$ factor is due to the aforementioned integrality gap while the second factor~$O(\log m)$ is due to the use of the potential function. To get a tight competitive ratio of $O(\log m)$, we avoid the integrality gap by comparing to a \emph{fractional} solution to \eqref{eq_lpstochconfig}, and we use a slightly different maximal inequality for the regime where the mean of the sums is larger than the increments by at most a $O(\log m)$-factor. The particular maximal inequality is the following, which we prove in \Cref{app_prelim}.



\begin{restatable}{lem}{lemmaxboundlogm}\label{lem_max_bound_logm}
	Let~$S_1, \ldots, S_m$ be sums of independent random variables bounded in~$[0,\tau]$ such that~$\EE[S_i] \leq O(\log m) \tau$ for all~$1 \leq i \leq m$. Then,~$\EE[\max_i S_i] \leq O(\log m) \tau$. 
\end{restatable}


Now we give our main guarantee for \Cref{alg_on_stoch_config}, which implies the desired $O(\log m)$-competitive online algorithm for configuration balancing with stochastic requests.

\begin{lem}\label{thm_cb_on}\label{lem_on_config_runtime}
	
	Suppose the minimizing configuration in Line \ref{alg_on_config_argmin} can be found in polynomial time. Then \Cref{alg_on_stoch_config} runs in polynomial time. Further, \Cref{alg_on_stoch_config} is a deterministic, non-adaptive algorithm that correctly solves the subproblem of \Cref{lem_guess_and_double} for $\alpha = O(\log m)$.
\end{lem}

\begin{proof}
	The running time of \Cref{alg_on_stoch_config} is clear. It remains to show that the algorithm creates expected makespan at most $O(\log m) \lambda$ or correctly certifies $\EE[\Opt] > \lambda$ if it terminates prematurely.
	
	We first show the former. By definition, upon termination of the algorithm after, say $n'$ requests, the expected truncated load on each resource and total expected exceptional load are both at most $O(\log m) \cdot \lambda$. Let the configuration choices $c_j$ be with respect to our algorithm. We can split the makespan into the contributions by the truncated and exceptional parts.
	\[
	\EE \bigg[\max_{1 \leq i \leq m } \sum_{j=1}^{n'} X_{ij}(c_j) \bigg] \leq \EE \bigg[\max_{1 \leq i \leq m} \sum_{j=1}^{n'} X_{ij}^T(c_j) \bigg] + \EE \bigg[\max_{1 \leq i \leq m} \sum_{j=1}^{n'} X_{ij}^E(c_j) \bigg]. 
	\]
	Each truncated part is bounded in $[0, 2\lambda]$ and each resource has expected truncated load at most $O(\log m) \lambda$. By \Cref{lem_max_bound_logm}, we can bound the contribution of the truncated parts by $\EE \big[\max_{1 \leq i \leq m} \sum_{j=1}^{n'} X_{ij}^T(c_j) \big] = O(\log m) \lambda$.
	For the exceptional parts, applying \eqref{eq_exceptional_proxy} gives
	\[
	\EE\bigg[\max_{1 \leq i \leq m} \sum_{j=1}^n X_{ij}^E(c_j) \bigg] \leq 
	\EE\bigg[\sum_{j=1}^n \max_{1 \leq i \leq m} X_{ij}^E(c_j) \bigg] \leq O(\log m) \lambda.
	\]
	Combining both bounds gives that the expected makespan is at most $O(\log m) \lambda$, as required.
	
	It remains to show that if $\EE[\Opt] \leq \lambda$, then the algorithm successfully assigns all requests. We do so by bounding the increase in the potential function. Note that if $\EE[\Opt] \leq \lambda$, then \eqref{eq_lpstochconfig} is feasible for our choice of $\tau = 2 \lambda$ by \Cref{lem_lpc_feasible}. Thus, let~$(y_{cj})$ be a feasible solution to~\eqref{eq_lpstochconfig}. For simplicity, let~$x_{0j}(c) := \EE[\max_{1\leq i \leq m} X_{ij}^E(c)]$ be the exceptional part of configuration~$c$ and~$x_{ij}(c) := \EE[X_{ij}^T(c)]$ its truncated part on resource $i$. 
	
	For each request $j$, as \Cref{alg_on_stoch_config} chooses a configuration~$c_j$ minimizing the increase in~$\Phi$, 
	\begin{align}
	    \phi(\load_{j-1} + x_j(c_j)) - \phi(\load_{j-1}) & \leq \phi(\load_{j-1} + x_j(c)) - \phi(\load_{j-1}) \notag 
	\shortintertext{for all configurations~$c \in [q_j]$. As~$\sum_{c=1}^{q_j} y_{cj} = 1$ by feasibility of~$(y_{cj})$, this implies}
	    \phi(\load_{j-1} + x_j(c_j)) - \phi(\load_{j-1}) & \leq \sum_{c=1}^{q_j} y_{cj} \phi(\load_{j-1} + x_{j}(c)) - \phi(\load_{j-1}). \label{eq_greedy_increase} 
	\shortintertext{\hspace*{\parindent}We recall that~$\load_0 = 0 \in \RR^{m+1}$. We bound the increase in potential incurred by \Cref{alg_on_stoch_config}:}
		\phi(\load_n) - \phi(\load_0) 
		& = \sum_{j=1}^n \sum_{i=0}^m (3/2)^{\load_{ij-1}/\tau} \Big( (3/2)^{x_{ij}(c_j)/\tau} - 1 \Big) \notag \\ 
		& \leq \sum_{j=1}^n \sum_{c=1}^{q_j} y_{cj} 
		\sum_{i=0}^m  (3/2)^{\load_{ij-1}/\tau} \Big( (3/2)^{x_{ij}(c)/\tau} - 1 \Big) \notag \\ 
		&  \leq 
		\sum_{i=0}^m (3/2)^{\load_{in}/\tau} \sum_{j=1}^n \sum_{c=1}^{q_j} y_{cj} \Big( (3/2)^{x_{ij}(c)/\tau} -1 \Big), \notag
	\shortintertext{where the first inequality holds due to~\eqref{eq_greedy_increase}, and the second inequality holds because the load on machine~$i$ only increases over time. 
			Standard estimates of $e^z$ give $(3/2)^z - 1 \leq (1/2) z$ for~$z \in [0,1]$. As~$y_{cj}$ is feasible, we know that~$x_{ij}(c) \leq \tau$ for all~$c$ with~$y_{cj} > 0$. Hence, }
		\phi(\load_n) - \phi(\load_0)  & \leq \sum_{i=0}^m (3/2)^{\load_{in}/\tau} \sum_{j=1}^n \sum_{c=1}^{q_j} y_{cj} \frac{x_{ij}(c)}{2\tau} \notag
	\shortintertext{Using that~$y_{cj}$ is feasible and satisfies $\sum_{j=1}^n \sum_{c=1}^{q_j} x_{ij}(c) \cdot y_{cj}  \leq \tau$ for resource~$i = 0$ by the exceptional constraint and for all resources $i= 1, \ldots, m$ by the truncated constraints, we get} 
		\phi(\load_n) - \phi(\load_0) & \leq (1/2) \sum_{i=0}^m (3/2)^{\load_{in}/\tau} = (1/2) \phi(\load_n). \notag
	\end{align}
	After rearranging, we have $\phi(\load_n) \leq 2 \phi(\load_0) = 2(m+1)$ by choice of~$\load_0$. Taking logarithms and using that~$\log_{3/2}(z)$ is monotonically increasing, we conclude that
	$	\max_{0 \leq i \leq m} \load_{in} \leq \log_{3/2} (2m + 2) \tau. $
	Note that we chose~$\ell = \log_{3/2} (2m + 2)$ implying that \Cref{alg_on_stoch_config} never fails if $\EE[\Opt] \leq \lambda$. 
\end{proof}

\subsection{Proof of \Cref{thm_config_balancing}}


\begin{proof}[Proof of \Cref{thm_config_balancing}]
    Note that the configurations are given explicitly, implying that both, \Cref{alg_off_config_stoch,alg_on_stoch_config}, run in polynomial time.
    Hence, \Cref{lem_off_config_stoch} gives the $O(\frac{\log m}{\log \log m})$-approximate offline algorithm and Lemmas \ref{lem_guess_and_double} and \ref{thm_cb_on} give the $O(\log m)$-approximate online algorithm. 
\end{proof}

\section{Unrelated Load Balancing and Virtual Circuit Routing}

In this section, we apply our algorithms for configuration balancing to stochastic load balancing on unrelated machines (\Cref{thm_unrel_lb_off}) as well as stochastic virtual circuit routing (\Cref{thm_routing}).

\subsection{Unrelated Load Balancing with Stochastic Jobs} 
We recall that in load balancing on unrelated machines with stochastic jobs, we have $m$ machines and $n$ jobs such that the size of job $j$ on machine $i$ is a random variable $X_{ij}$. These $X_{ij}$s are independent across jobs. 
This is a special case of configuration balancing with stochastic requests by taking $m$ resources (corresponding to the $m$ machines) and $n$ requests such that each request $j$ has $m$ possible configurations, one for each machine choice job $j$ has. Precisely, we define the configurations $c = 1, \ldots, m$ for job $j$ by setting
\begin{equation}\label{eq_LBtoCB}
	X_{ij}(c) = 
	\begin{cases} 
		X_{cj}  & \textup{ if } i = c, \\ 
		0 & \textup{ otherwise}.
	\end{cases}
\end{equation}

\begin{proof}[Proof of \Cref{thm_unrel_lb_off}] 
    Each request has $m$ possible configurations, so the total size of the resulting configuration balancing instance is polynomial. Thus, we may assume the configurations are given explicitly.
	Hence, \Cref{thm_config_off} immediately gives a  \emph{randomized} $O(\frac{\log m}{\log \log m})$-approximation offline and $O(\log m)$-approximation online for load balancing on unrelated machines with stochastic jobs.
	
	However, to obtain a deterministic offline algorithm, we can de-randomize \Cref{alg_off_config_stoch} for this special case because here, \eqref{eq_lpstochconfig} is equivalent to the generalized assignment LP considered by Shmoys and Tardos, which has a constant-factor rounding algorithm \cite{DBLP:journals/mp/ShmoysT93}.
\end{proof}

\subsection{Routing with Stochastic Demands}

Virtual circuit routing is another special case of configuration balancing: Given any instance of the former, which consists of a directed graph with $m$ edges and $n$ source-sink pairs, the resulting configuration balancing instance has $m$ resources, one for each edge, and $n$ requests, one for each source-sink pair, such that there is one configuration per source-sink path in the graph.

Note that unlike load balancing, where a request has at most $m$ configurations, for routing, a request can have exponentially many configurations. Thus, to obtain polynomial-time algorithms for routing, we cannot explicitly represent all configurations. In particular, to prove \Cref{thm_route_off} in the offline setting, we need to show how to solve \eqref{eq_lpstochconfig} efficiently in the special case of routing, and to prove \Cref{thm_route_off} in the online setting, we need to show how to find the configuration (i.e. the path) that minimizes the increase in the potential $\Phi$. 

\subsubsection{Offline Routing: Solving \eqref{eq_lpstochconfig}}

We first re-write \eqref{eq_lpstochconfig} for routing. Recall that~$\tau$ is the truncation threshold. For a request~$j$, let~$E_j$ be the set of all edges with~$\EE[X_{ej}] \leq \tau$ and let $\mathcal P_j$ be the collection of all $s_j$-$t_j$ paths in the auxiliary graph $G_j = (V,E_j)$. For each request~$j$ and path~$P \in \mathcal P_j$, the variable~$y_{Pj}$ denotes the decision to route request~$j$ along path $P$. Thus, we want to solve the following path assignment~LP.


\begin{equation}\tag{$\lppath$}\label{eq_lppath}
	\begin{array}{rll}
		\sum_{P \in \mathcal P_j} y_{Pj} &  = 1 &\quad\forall~j\in[n] \\
		\sum_{j=1}^n \sum_{P \in \mathcal P_j} \EE[X_{ej}^T] \cdot y_{Pj}  &  \leq \tau &\quad\forall~e \in E \\
		\sum_{j=1}^n \sum_{P \in \mathcal P_j} \EE[\max_{e \in P} X_{ej}^E] \cdot y_{Pj} & \leq \tau \\
		y_{Pj} & \geq 0 & \quad \forall~j\in[n], ~ P \in \mathcal P_j 
	\end{array} 
\end{equation}
 
To see that \eqref{eq_lppath} is equivalent to \eqref{eq_lpstochconfig}%
, note that the pruning constraints of the latter ensure that any configuration/path $P$ with $\EE \big[ \max_{e \in P} X_{ej} \big] > \tau$ will not be selected. Re-writing gives 
\(\EE \big[ \max_{e \in P} X_{ej} \big] = \EE \big[ \max_{e \in P} \frac{1}{c_e} X_j \big]= \max_{e \in P} \EE[X_{ej}].\)
Thus, the pruning constraint ensures that no edge with $\EE[X_{ej}] > \tau$ will be selected. This is exactly encoded by the set of feasible paths $\mathcal{P}_j$.

Note that \eqref{eq_lppath} has an exponential number of variables. For classical congestion minimization problems, the path-assignment LP formulation is equivalent to a flow formulation \cite{LeightonRS98}, which can be solved optimally in polynomial time using results about flows in networks. In our case, however, we additionally have the third constraint (the exceptional load constraint) in the LP, which does not allow for a straight-forward equivalent flow formulation. Therefore, we use LP duality in order to solve it efficiently: We give a separation oracle for its dual LP. 
For obtaining the dual, we add the trivial objective to maximize $0^T y$ to \eqref{eq_lppath}. Hence, the dual of \eqref{eq_lppath} is 

\begin{equation}\tag{$\lppathd$}\label{eq_lppathd}
	\begin{array}{rll}
	    \min \sum_j a_j + \sum_e b_e \cdot \tau + c \cdot \tau &\\
		s.t. \quad a_j + \sum_{e \in P} b_e \cdot \EE[X_{ej}^T] + c \cdot \EE[\max_{e \in P} X_{ej}^T] & \geq 0 & \quad\forall j \in [n], ~P \in\mathcal P_j\\
		b_e & \geq 0 &\quad\forall~e \in E \\
		c & \geq 0.
	\end{array} 
\end{equation}

For solving \eqref{eq_lppathd}, consider a request~$j$ and a path~$P \in \mathcal P_j$. The expected exceptional part of routing~$j$ along~$P$ is $\EE \big[\max_{e \in P} X_{ej} ^E \big] =  \max_{e \in P} \EE \big[\big({X_j}/{c_e}\big)^E\big]$, which is the expected exceptional part of the smallest-capacity edge~$\bar e_j$ along the path. Given this edge~$\bar e_j$, a particular choice of the dual variables is feasible for the first constraint of \eqref{eq_lppathd} if and only if
\[\min_{P \in \mathcal P_j \, : \, \min_{e \in P} c_e \geq c_{\bar e_j} } \bigg( \sum_{e \in P} b_e \cdot \EE\big[X_{ej}^T\big] + c \cdot \EE\big[\max_{e \in P} X_{ej}^T\big] \bigg) \geq -a_j \quad\forall~ j.\] 

Hence, for each request~$j$, it remains to find a path~$P \in \mathcal P_j$ that is minimal w.r.t. edge weights $b_e \cdot \mathbb{E}[X_{ej}^T]$. By letting every edge~$\bar e$ be the smallest-capacity edge and removing any smaller-capacity edges from the graph, this becomes a shortest~$s_j$-$t_j$ path problem. 

\begin{algorithm}
    \DontPrintSemicolon
    \caption{Separation oracle for \eqref{eq_lppathd}}
    \label{alg_route_off} 
    $(a,b,c) \leftarrow$ current solution of \eqref{eq_lppathd} \;
    \uIf{$b_e < 0$ for some $e \in E$ {\normalfont \bfseries or }$c < 0$} {
        return the violated non-negativity constraint\;
    }
    \Else{
        \For{$j$ {\normalfont \bfseries and }edge $\bar{e} \in E_j$}{
            $E_{\bar e}\leftarrow \{e \in E_j: c_e \geq  c_{\bar{e}} \}$ \;
            $G_{\bar{e}} \leftarrow$ $(V, E_{\bar e}) $ \;
            $P_{j\bar e} \leftarrow$ shortest $s_j$-$t_j$ path in $G_{\bar{e}}$ w.r.t. edge weights $b_e \cdot \mathbb{E}[X_{ej}^T]$ \;
            \If{$\sum_{e \in P_{j\bar{e}}} b_e \cdot \EE[X_{ej}^T] + c \cdot \EE[\max_{e \in P_{j\bar{e}}} X_{ej}^T] < - a_j$} {
                return the separating hyperplane $\sum_{e \in P_{j\bar{e}}} b_e \cdot \EE[X_{ej}^T] + c \cdot \EE[\max_{e \in P_{j\bar{e}}} X_{ej}^T] \geq - a_j$\;
            }
        }
        return ``feasible''\;
    }
\end{algorithm}


\begin{lem}\label{lem_routing_dual}
    \Cref{alg_route_off} is a polynomial time separation oracle for \eqref{eq_lppathd}. 
\end{lem}

\begin{proof}
    Since shortest paths can be found in polynomial time \cite{Bellman58}, \Cref{alg_route_off} indeed runs in polynomial time. Further, if there is an edge~$e \in E$ with~$b_e < 0$ or if~$c<0$, then \Cref{alg_route_off} correctly outputs the violated constraint. 
    
    It remains to consider the case when~$b, c \geq 0$. In this case, we need to show for each request~$j$ that we find some $\bar{e} \in E$ such that $P_{j \bar{e}}$ achieves 
    $\min_{P \in \mathcal P_j} \big( \sum_{e \in P} b_e \cdot \EE[X_{ej}^T] + c \cdot \EE[\max_{e \in P} X_{ej}^T] \big)$. 
    
    Consider any request $j$ such that the minimum is achieved by some $s_j$-$t_j$ path $P^*$. Let $e^*$ be the smallest-capacity edge along $P^*$. We claim for the correct guess $\bar{e} = e^*$, $P_{j \bar{e}}$ achieves the minimum. To see this, observe that $P^*$ is a $s_j$-$t_j$ path in the graph $G_{\bar{e}}$, so the algorithm will choose~$P_{j \bar{e}}$ with $\sum_{e \in P_{j\bar{e}}} b_e \cdot \EE[X_{ej}^T] \leq \sum_{e \in P^*} b_e \cdot \EE[X_{ej}^T]$ by definition of the edge weights. Further, we have $c \cdot \EE[\max_{e \in P_{j\bar{e}}} X_{ej}^T] \leq c \cdot \EE[\max_{e \in P^*} X_{ej}^T]$, because the latter maximum is achieved by edge $\bar{e}$, and by definition of the residual graph, $P_{j \bar{e}}$ cannot use any edges with smaller capacity than $\bar{e}$. Combining these two bounds shows that $P_{j \bar{e}}$ achieves the minimum. This in turn implies that \Cref{alg_route_off} returns a constraint violated by~$(a,b,c)$ if such a constraint exists. 
\end{proof}

\subsubsection{Online Routing: Minimizing Increase in $\Phi$}

For online virtual circuit routing, we assume that a  sequence of source sink-pairs $(s_j,t_j)$ for $j = 1, \dots, n$ arrive online. To implement \Cref{alg_on_stoch_config} efficiently, given a load vector $\load = (\load_0, \dots, \load_m)$ and source-sink pair $(s_j,t_j)$ with random demand $X_j \geq 0$, we need to choose a $s_j$-$t_j$ path in $G$ that minimizes the increase in $\Phi$ with respect to some fixed truncation threshold $\tau$. Recall that we index the edges of $G$ by $1, \dots, m$, while $L_0$ is the load of an additional, artificial resource that captures the total expected exceptional load.

As in the offline setting, the expected exceptional part of the configuration corresponding to choosing a particular~$s_j$-$t_j$ path is the expected exceptional part of the smallest-capacity edge along the path. Thus, the increase in potential due to choosing a path~$P$ is
\[
    \big( (3/2)^{ (\load_{0} + ( \max_{e \in P} \EE [X_{ej}^E]) / \tau } - (3/2)^{ \load_{0}  / \tau    }\big) + 
    \sum_{e \in P}  \big( (3/2)^{ ( \load_{e} + \EE[X_{ej}^T])    / \tau } - (3/2)^{  \load_{e}    / \tau }\big).
\]
The first term is the increase due to the exceptional part (the smallest-capacity edge along $P$), and the remaining terms are the per-edge contributions due to the truncated parts. Analogously to the previous section, we only consider edges in $E_j = \{e \in E: \EE[X_{ej}]\leq \tau\}$, 
guess the smallest-capacity edge, and solve a shortest $s_j$-$t_j$ path problem to find the minimizing path; see 
\Cref{alg_route_on}. 

\begin{algorithm}
    \DontPrintSemicolon
    \caption{Minimizing increase in $\Phi$ for virtual circuit routing}
    \label{alg_route_on} 
    $L\leftarrow$ current load vector \;
    \For{$\bar{e} \in E_j$}{
        $E_{\bar e}\leftarrow \{e \in E_j: c_e \geq  c_{\bar{e}} \}$ \;
        $G_{\bar{e}} \leftarrow$ $(V, E_{\bar e}) $ \;
        $P_{j\bar e} \leftarrow$ shortest $s_j$-$t_j$ path in $G_{\bar{e}}$ w.r.t. edge weights $\big( (3/2)^{ ( \load_{e} + \EE[X_{ej}^T])    / \tau } - (3/2)^{  \load_{e}    / \tau }\big)$  \;
    }
    $P_{j} \leftarrow \arg \min_{P_{j\bar e}}$ increase in $\Phi$ \;
\end{algorithm}


\begin{lem}\label{lem_route_minincrease}
    Given a request~$j$ and load vector $L$, \Cref{alg_route_on} returns in polynomial time a path~$P_j$ that minimizes the increase in~$\Phi$. 
\end{lem}

\begin{proof}
    It is clear that \Cref{alg_route_on} runs in polynomial time. To see that it also finds a path that minimizes the increase in $\Phi$, let $P_j^*$ be such an optimal path with smallest-capacity edge~$e^*$. For the correct guess~$\bar{e} = e^*$,~$P_j^*$ is a $s_j$-$t_j$ path in the graph~$G_{\bar e}$. Thus, \Cref{alg_route_on} chooses $P_{j\bar e}$ such that the per-edge contributions due to the truncated parts of $P_{j\bar e}$ are at most those due to $P_j^*$ by definition of the edge weights. Further, the exceptional part of $P_{j\bar e}$ is at most that of $P_j^*$, because $P_{j \bar e}$ does not use any edges with capacity smaller than~$c_{\bar e}$ by definition of the graph~$G_{\bar e}$. We conclude that~$P_{j\bar e}$ is also a~$s_j$-$t_j$ path that minimizes the increase in $\Phi$. 
\end{proof}

\paragraph{Proof of \Cref{thm_routing}}

\begin{proof}[Proof of \Cref{thm_routing}] 

For offline virtual circuit routing, \Cref{lem_routing_dual} guarantees that \eqref{eq_lppath} can be solved optimally in polynomial time by LP duality. Thus, \Cref{lem_off_config_stoch} implies that \Cref{alg_off_config_stoch} runs in polynomial time and achieves a maximum congestion of $O\big(\frac{\log m}{\log \log m}\big) \EE[\Opt]$. 

For 
the online problem, \Cref{lem_route_minincrease} guarantees that, for each request~$j$, a path~$P_j$ that minimizes the increase in the potential function~$\Phi$ is found in polynomial time. Thus, \Cref{lem_off_config_stoch} implies that \Cref{alg_on_stoch_config} runs in polynomial time and guarantees a maximum congestion of~$O(\log m)\EE[\Opt]$. 
\end{proof}

\section{Load Balancing on Related Machines}\label{sec_related}
In this section, we improve on \Cref{thm_unrel_lb_off} in the special case of related machines, where each machine $i$ has a speed parameter $s_i > 0$ and each job $j$ an independent size $X_j$ such that $X_{ij} = \frac{X_j}{s_i}$. Recall that we gave a non-adaptive $O\big(\frac{\log m}{\log \log m}\big)$-approximation for unrelated machines. However, the adaptivity gap is $\Omega\big(\frac{\log m}{\log \log m}\big)$ even for load balancing on identical machines where every machine has the same speed. Thus, to improve on \Cref{thm_unrel_lb_off}, we need to use adaptivity.

The starting point of our improved algorithms is the same non-adaptive assignment for unrelated-machine load balancing. 
However, instead of non-adaptively assigning a job $j$ to the specified machine $i$, we adaptively assign $j$ to the least loaded machine with similar speed to $i$. In the first part, we formalize this idea 
and then we focus on offline and online load balancing on related machines. 

\subsection{Machine Smoothing}\label{subsec_machine_smoothing}

In this section, we define a notion of \emph{smoothed machines}. We show that by losing a constant factor in the approximation ratio, we may assume that the machines are partitioned into at most $O(\log m)$ groups such that machines within a group have the same speed and the size of the groups shrinks geometrically. Thus, by ``machines with similar speed to $i$,'' we mean machines in the same group. 

Formally, we would like to transform an instance~$\mathcal I$ of load balancing on~$m$ related machines with stochastic jobs into an instance~$\mathcal I_s$ with so-called ``smoothed machines'' and the same set of jobs with the following three properties:  
\begin{enumerate}[label=(\roman*)]
        \item The machines are partitioned into $m' = O(\log m)$ groups such that group $k$ consists of $m_k$ machines with speed exactly $s_k$ such that $s_1 < s_2 < \dots < s_{m'}$.\label{smoothed_prop_speeds}
        \item For all groups $1 \leq k < m'$, we have $m_k \geq \frac{3}{2} m_{k+1}$. \label{smoothed_prop_groups}
        \item $\Opt (\mathcal I_s) = O( \Opt (\mathcal I))$.\label{smoothed_prop_cost}
\end{enumerate}

To this end, we suitably decrease machine speeds and delete machines from the original instance~$\mathcal I$. Our algorithm is the following.

\begin{algorithm}
    \DontPrintSemicolon
    \caption{Machine Smoothing}
    \label{alg_mach_smooth} 
    $s_{\max} \leftarrow \max_{i} s_{i}$ \;
    \For{$i =1$ to $m$}{
        $s_i \leftarrow s_i / s_{\max}$ \;
        \uIf{$s_i \leq \frac1m$}
        {
           \nl delete machine~$i$ \label{alg_ms_delete_slow}\;
        }
        \Else{
        \nl $s_i \leftarrow 2^{\lfloor \log s_i \rfloor}$ 
        \label{alg_ms_round_down}\;
        }
    }
    partition machines by speeds such that group~$k$ has~$m_k$ machines of speed~$s_k$\;
    index the groups in order of increasing speed \; 
    \For{$k=1$ to $\lceil \log m \rceil$}{
        \If{ $m_k < \frac{3}{2} m_{k+1}$}{
            \nl delete group k \label{alg_ms_delete_small}\;
        }
    } 
\end{algorithm}

We show that this algorithm creates the desired smoothed machines instance.   

\begin{restatable}{lem}{lemrelsmooth}\label{lem_rel_smooth}
	Given an instance~$\mathcal I$ of load balancing with~$m$ related machines and stochastic jobs, \Cref{alg_mach_smooth} efficiently computes an instance~$\mathcal I_s$ of smoothed machines with the same set of jobs satisfying Properties~\ref{smoothed_prop_speeds} to~\ref{smoothed_prop_cost}. 
\end{restatable}
\begin{proof}[Proof sketch (full proof in \Cref{sec_appendix_related})]
It is clear that the algorithm is efficient and outputs $\mathcal{I}_s$ satisfying~\ref{smoothed_prop_speeds} and~\ref{smoothed_prop_groups}. For showing~\ref{smoothed_prop_cost}, we analyze the increase of the cost of \Opt{} due to each step. For Step~\ref{alg_ms_delete_slow}, we schedule the jobs assigned by \Opt{} to deleted machines on the fastest machine, increasing its load by at most~$(m-1) \frac{1}{m} \Opt$. Step~\ref{alg_ms_round_down} increases \Opt{} by at most a factor~$2$. For Step~\ref{alg_ms_delete_small}, we schedule all jobs assigned by \Opt{} to a deleted machine on machines of the next faster remaining group, following a fixed mapping of the deleted machines to the remaining machines. Because of~\ref{smoothed_prop_groups}, we can bound the increase in load on each remaining machine by~$O(\Opt)$. 
\end{proof}

To summarize, by losing a constant-factor in our final approximation ratio, we may assume we are working with an instance of smoothed machines. Looking ahead, if our non-adaptive policy assigns job $j$ to machine $i$, then we instead adaptively assign $j$ to the least-loaded machine in the group containing machine $i$. We will use the properties of smoothed machines to show that this leads to a $O(1)$-approximation offline and $O(\log \log m)$-approximation online.

A similar idea for machine smoothing has been employed by Im et al.~\cite{ImKPS18} for deterministic load balancing on related machines. In their approach, they ensure that the \emph{total processing power} of the machines in a group decreases geometrically rather than the number of machines.

\subsection{Offline Setting}\label{sec_off_rel}

In this section, we give our $O(1)$-approximate adaptive policy for load balancing on related machines with stochastic jobs. Our algorithm has three parts: machine smoothing (\Cref{alg_mach_smooth}), non-adaptive assignment (\Cref{alg_off_config_stoch}), and adaptively turning a job-to-machine assignment into a coarser job-to-group assignment. Precisely, our algorithm is the following.

\begin{algorithm}
	\DontPrintSemicolon
	\caption{Offline Related Load Balancing}
	\label{alg_rel_off}
    smooth machines with \Cref{alg_mach_smooth} \;
    create configuration balancing instance as in \eqref{eq_LBtoCB}\;
    obtain assignment of jobs to machines by \Cref{alg_off_config_stoch}\;
    \For{$j \in J$}{
        $i \leftarrow$ machine of~$j$ \;
        schedule $j$ on least loaded machine~$i'$ with $s_{i} = s_{i'}$ \;
    }
\end{algorithm}

Note that as in the case of unrelated machine load balancing (\Cref{thm_unrel_lb_off}), we can derandomize this algorithm by employing the GAP LP rounding algorithm by~\cite{DBLP:journals/mp/ShmoysT93}. 




We show that this algorithm gives the desired $O(1)$-approximation. Note that our previous analysis of \Cref{alg_off_config_stoch} gave a $O(\frac{\log m}{\log \log m})$-approximation. We improve on this using the properties of smoothed machines and our adaptive decisions. We give a stronger maximal inequality using the fact that group sizes are geometrically increasing. See \Cref{app_prelim} for proof of the following.


\begin{restatable}{lem}{lemmaxboundgeo}\label{lem_max_bound_geo}
    Let~$c_1,\ldots, c_m \in \mathbb N_{\geq 1}$ be constants such that~$c_i \geq \frac32 c_{i+1}$ for all~$1 \leq i \leq m$.  
    Let $S_1,\ldots,S_m$ be sums of independent random variables bounded in $[0,\tau]$ such that~$\EE[S_i] \leq c_i \tau$ for all~$1 \leq i \leq m$. Then, $\EE\big[\max_i \frac{S_i}{c_i}\big] \leq O(\tau)$. 
\end{restatable}

Now we analyze \Cref{alg_rel_off}.




\begin{lem}\label{lem_rel_lb_off}
    For offline load balancing on related machines with stochastic jobs, \Cref{alg_rel_off} efficiently 
    either outputs an adaptive policy with expected makespan $O(1)$ or certifies~$\EE[\Opt] > 1$. 
\end{lem}
\begin{proof}
	It is clear that the algorithm runs in polynomial time. By the contrapositive of \Cref{lem_lpc_feasible}, if \eqref{eq_lpstochconfig} is infeasible for $\tau = 2$, then we correctly certify $\EE[\Opt] > 1$. 
	Thus, it remains to consider the case where the LP is feasible. 
	
	In this case, we obtain a job-to-machine assignment with expected truncated load at most $2$ on every machine and expected exceptional load at most $2$. Summing up the truncated loads within each group, we have that group $k$ has expected truncated load at most $2m_k$. Again, we split the makespan of our policy into truncated and exceptional parts
	\[
	\EE\bigg[ \max_{1 \leq i \leq m}  \sum_{j \rightarrow i} X_{ij} \bigg] \leq \EE\bigg[ \max_{1 \leq i \leq m}  \sum_{j \rightarrow i} X_{ij}^T \bigg] + \EE\bigg[ \max_{1 \leq i \leq m}  \sum_{j \rightarrow i} X_{ij}^E \bigg],
	\]
	where the events $j \rightarrow i$ are with respect to our final adaptive assignment. We can upper bound the contribution of the exceptional parts (the latter term) by $2$ using \eqref{eq_exceptional_proxy}. It remains to bound the contribution of truncated parts. We do so by considering the makespan on each \emph{group}. For group~$k$, we let $j \rightarrow k$ denote the event that we non-adaptively assign job~$j$ to a machine in group~$k$, and $X_{kj}$ be the size of job~$j$ on any machine in group $k$ (recall that they all have the same speed).  Then, 
	\[\EE\bigg[ \max_{1 \leq i \leq m}  \sum_{j \rightarrow i} X_{ij}^T \bigg] = \EE\bigg[\max_k  \max_{i \in G_k}  \sum_{j \rightarrow i} X_{ij}^T \bigg],\]
	where $G_k$ is the collection of all machines in group $k$. Similar to the analysis of list scheduling by~\cite{DBLP:journals/siamam/Graham69}, i.e, by an averaging argument, we obtain
	\begin{equation}\label{eq_vol_off}
	\max_{i \in G_k}  \sum_{j \rightarrow i} X_{ij}^T \leq \frac{1}{m_k} \sum_{j \rightarrow k} X_{kj}^T + \max_{j \rightarrow k} X_{kj}^T \leq \frac{1}{m_k} \sum_{j \rightarrow k} X_{kj}^T + 2.
	\end{equation}
	The expected truncated load on group $k$ is at most $2 m_k$, and we have $m_k \geq \frac{3}{2} m_{k+1}$ for all $k$ by the properties of smoothed machines. 
	\Cref{lem_max_bound_geo} bounds the expected maximum of \eqref{eq_vol_off} for all $k$ by $O(1)$. This upper bounds the expected contribution of the truncated parts by $O(1)$, as required.
\end{proof}

\subsection{Online Load Balancing on Related Machines}\label{sec_on_rel}

In this section, we apply the same framework to the online setting. As main difference, we compute the non-adaptive assignment online using \Cref{alg_on_stoch_config}. We must be careful, because our online configuration balancing algorithm loses a logarithmic factor in the number of resources, so to obtain a $O(\log \log m)$-approximation, we aggregate each group (in the smoothed-machines instance) as a single resource. Thus, our configuration balancing instance will have only $O(\log m)$ resources.

We first describe how to construct the configuration balancing instance. Suppose we have smoothed machines with $m'$ groups. We define  $m' + 1$ resources indexed $0, \dots, m'$ such that the~$0$th resource is a virtual resource collecting exceptional parts, and the resources $1, \dots m'$ index the groups of smoothed machines and collect the respective truncated parts. Each job $j$ has $m'$ configurations $c = 1, \dots, m'$ (one per job-to-group assignment) defined by
\begin{equation}\label{eq_RLB_CB}
	X_{kj}(c) = 
	\begin{cases} 
		\EE\big[ X_{kj}^E \big] & \textup{ if } k = 0, \\
		\frac{1}{m_k} \EE\big[ X_{kj}^T\big] & \textup{ if } k = c, \\
		0 & \textup{ otherwise},
	\end{cases}
\end{equation}
for a fixed truncation threshold $\tau$. Note that these configurations are  deterministic. Intuitively, this definition captures the fact that we will average all jobs assigned to group $k$ over the $m_k$ machines in the group. Now we give our algorithm for the online setting.

\begin{algorithm}
	\DontPrintSemicolon
	\caption{Online Related Load Balancing}
	\label{alg_rel_on}
	smooth machines with \Cref{alg_mach_smooth}\;
	$\lambda \leftarrow$ current guess of~$\EE[\Opt]$\;
	$\tau \leftarrow 2 \lambda$ truncation threshold\;
	\Upon{arrival of request $j$}{
    	construct configurations as in \eqref{eq_RLB_CB}\; 
    	$k \leftarrow$ configuration chosen by \Cref{alg_on_stoch_config}\;
    	schedule $j$ on least-loaded machine in group $k$ \;
	}
\end{algorithm}

\begin{lem}\label{lem_rel_lb_on}
	For online load balancing on related machines with stochastic jobs, \Cref{alg_rel_on} runs in polynomial time and correctly solves the subproblem of \Cref{lem_guess_and_double} for $\alpha = O(\log \log m)$. 
\end{lem}
\begin{proof}
	It is clear that the algorithm runs in polynomial time.
	
	Let $\Opt_{S}$ be the optimal makespan of the smoothed load balancing instance. We first claim that the resulting configuration balancing instance has optimal makespan at most $O(\EE[\Opt_{S}])$ for any truncation threshold $\tau \geq 2 \EE[\Opt_{S}]$. To see this, observe that \eqref{eq_lpstochconfig} for the smoothed load balancing instance is feasible for~$\tau$. Then, \eqref{eq_lpstochconfig} for the deterministic configuration balancing instance is obtained from this LP by aggregating the truncated constraints for all machines in the same group and dividing by $m_k$. Thus, this latter LP is also feasible for the same threshold, and it admits a constant-factor approximation by Shmoys-Tardos \cite{DBLP:journals/mp/ShmoysT93}. 	Thus, $\Opt_D$, the optimal solution of the deterministic configuration balancing instance defined by~\eqref{eq_RLB_CB}, is at most~$O(\EE[\Opt_S])$.
	
	
	Further, by the properties of smoothed machines, the resulting configuration balancing instance has $m' + 1 = O(\log m)$ resources. As argued in the proof of \Cref{thm_rel_lb_off}, the potential function guarantees that \Cref{alg_on_stoch_config} never creates makespan (for the configuration balancing instance) more than $O(\log \log m) \lambda$ if~$\Opt_D \leq \lambda$. Hence, by~\eqref{eq_RLB_CB}, the total expected exceptional load and average expected truncated load within each group are at most $O(\log \log m) \lambda$. We translate these bounds into bounds on the makespan for the load balancing instance.
	
	Similar to \Cref{lem_rel_lb_off}, we can upper bound the contribution of the exceptional parts by $O(\log \log m) \lambda$ using \eqref{eq_exceptional_proxy} and the truncated parts by $O(\log \log m) \lambda$ using \Cref{lem_max_bound_geo}. Thus, the expected makespan of the load balancing instance is at most $O(\log \log m) \lambda$, as required.
	
	Finally, we recall that if \Cref{alg_on_stoch_config} fails, then~$\Opt_D>\lambda$, which implies $\EE[\Opt_S] \geq \Omega(\lambda)$. 
\end{proof}

\paragraph{Proof of \Cref{thm_related}}
The proof follows immediately from \Cref{lem_rel_lb_off} and \Cref{lem_rel_lb_on}.

\section*{Conclusion}
We considered the
configuration balancing problem under uncertainty. In
contrast to the (often overly optimistic) clairvoyant settings and the
(often overly pessimistic) non-clairvoyant settings, we consider the
stochastic setting where each request~$j$ presents a set of
random vectors, and we need to (adaptively) pick one of these vectors, to minimize the \emph{expected} maximum load over the $m$
resources. We give logarithmic bounds for several general settings
(which are existentially tight), and a much better $O(1)$ offline and
$O(\log \log m)$ online bound for the related~machines
setting. Closing the gap for online related-machines load balancing remains an intriguing concrete open
problem. More generally, getting a better understanding of both
adaptive and non-adaptive algorithms for stochastic packing and
scheduling problems remains an exciting direction for
research. 

\newpage 

\appendix

\section{Inequalities and Bounds on Sums of Random Variables}\label{app_prelim}

In this section, we state some useful probabilistic inequalities. Further, we give the proofs of Lemmas~\ref{lem_max_bound}, \ref{lem_max_bound_logm}, and \ref{lem_max_bound_geo}.

\begin{lem}[Bernstein's inequality]\label{lem_bernstein}
    Let the random variables $X_1, \ldots, X_n$ be independent with $X_i - \EE[X_j] \leq b$ for each $1\leq j \leq n$. Let $S=\sum_{j=1}^n X_j$ and let $\sigma^2 = \sum_{j=1}^n \sigma_j^2$ be the variance of~$S$. Then, for any~$t >0$, \[
        \PP[S > \EE[S] + t] \leq \textup{exp} \bigg( - \frac{t^2}{ 2\sigma^2 (1 + bt/3\sigma^2)} \bigg).  \]
\end{lem}

\begin{lem}[Bennett's inequality  \cite{Bennett62}]\label{prop_bennett}
        Let $X_1,\dots, X_n$ be independent random variables with zero mean such that $X_j \leq a$ for all $j$. Let $S = \sum_j X_j$ and $\sigma^2 = \sum_j \mathbb{E}[X_j^2]$. Then, for all $t \geq 0$, 
        \[\mathbb{P}(S > t) \leq \exp\bigg( \frac{-\sigma^2}{a^2} \bigg(\bigg(1 + \frac{at}{\sigma^2}\bigg)\log\bigg(1 + \frac{at}{\sigma^2}\bigg) - \frac{at}{\sigma^2}\bigg)\bigg).\]
\end{lem}

\begin{lem}[Chernoff-Hoeffding type inequality; Theorem 1.1 in \cite{DubhashiP2009}]\label{lem_chernoff}
    Let $S= \sum_{j=1}^n X_j$ where $X_j$, $1 \leq j \leq n$, are independently distributed in $[0,1]$. If $t > 2e\EE[S]$, then 
    \[\PP[S > t] \leq 2^{-t}.\] 
\end{lem} 

Having these inequalities at hand, we are now ready to prove Lemmas~\ref{lem_max_bound}, \ref{lem_max_bound_logm}, and \ref{lem_max_bound_geo}, that bound the expected maximum of sums of independent variables.

\lemmaxbound*
\begin{proof}
    By re-scaling, it suffices to prove the lemma for $\tau = 1$. Consider one such sum, say $S = \sum_j X_j$, where the $X_j$s are independent and bounded in $[0, 1]$. We use Bennett's inequality (\Cref{prop_bennett}) to bound the upper tail of $S$. Applying Bennett's inequality to $S - \mathbb{E}[S]$ with $a = 1$ and $ \sigma^2 = \sum_j \mathbb{E}\big[(X_j - \mathbb{E}[X_j])^2\big] \leq \sum_j \mathbb{E}[X_j^2] \leq \sum_j \mathbb{E}[X_j] \leq 1$ gives for any $t \geq 0$:
    \begin{align*}
        \mathbb{P}[S > 1 + t] &\leq \mathbb{P}[S > \mathbb{E}[S]+t]\\
        &\leq \exp \bigg( - \big(\sigma^2 + t\big) \log\bigg(1 + \frac{t}{\sigma^2}\bigg) + t \bigg)\\
        &\leq \exp \big( - t \log(1 + t) + t \big) = \frac{e^t}{(1 + t)^t},
    \end{align*}
    where we use~$0\leq \sigma^2 \leq 1$ in the third inequality. 
    In particular, for $\ell = O\big(\frac{\log m}{\log \log m}\big)$ large enough, we have for any $t \geq 0$:
    \begin{equation*}
        \mathbb{P}[S > 1 + \ell + t] \leq \frac{e^{\ell + t}}{(1 + \ell + t)^{\ell + t}} \leq \frac{e^\ell}{\ell^\ell} \cdot \frac{e^t}{\ell^t} = O\bigg(\frac{1}{m}\bigg) \cdot e^{-t}.
    \end{equation*}
    This tail bound holds for all $S_1, \dots, S_m$. Union-bounding over all $m$ sums gives:
    \begin{align*}
        \mathbb{E} \big[\max_i S_i\big] &= \int_0^\infty \mathbb{P}\big[ \max_i S_i > t\big] \cdot dt\\
        &\leq (1 + \ell) + \int_0^\infty \mathbb{P}\big[\max_i S_i > 1+ \ell + t\big] \cdot dt\\
        &\leq (1 + \ell) + \sum_i \int_0^\infty \mathbb{P}[S_i > 1 + \ell + t] \cdot dt\\
        &= (1 + \ell) + m \cdot O\bigg(\frac{1}{m}\bigg) \int_0^\infty e^{-t} \cdot dt = O(\ell) = O \bigg(\frac{\log m}{\log \log m} \bigg).
    \end{align*}
\end{proof}

\lemmaxboundlogm*

\begin{proof}
    Let~$C \geq 0$ be a constant such that~$\EE[S_i] \leq C \tau \log m$ for all~$1 \leq i \leq m$ and fix~$t > 2e C \tau \log m$, where~$e$ is the Euler constant. Fixing~$i\in[m]$, we know that the random variables constituting~$S_i$ are independent and bounded by~$\tau$. Hence, using the Chernoff-Hoeffding bound in \Cref{lem_chernoff}, we have~$\PP[S_ i > t] < 2^{-t}$. Therefore,
    \begin{align*}
        \EE\Big[ \max_{1 \leq i \leq m} S_i \Big] \notag
        & = \int_{0}^\infty \PP\Big[ \max_{1 \leq i \leq m} S_i  > t  \Big] dt \\
        & \leq O(\log m)\tau + \int_{2eC \tau \log m}^\infty  \PP\Big[ \max_{1 \leq i \leq m} S_i  > t  \Big] dt \\
        & \leq O(\log m)\tau + \sum_{i=1}^m \int_{2eC \tau \log m}^\infty \PP\Big[ S_i  > t  \Big] dt \\
        & \leq O(\log m)\tau + m \int_{2 e C \tau \log m}^\infty 2^{-t} dt \\
        & = O(\log m) \tau, 
    \end{align*}
    where the third line uses a union bound over all~$i \in [m]$. 
\end{proof}

\lemmaxboundgeo*

\begin{proof}
     By re-scaling, we can assume~$\tau = 1$. Fix one sum, say~$S = \sum_j X_j$. The~$X_j$s are independent with~${X_j}/ c \leq \frac1c$ and $ {X_j}/{c} - \EE\big [ {X_j}/{c} \big] \leq \frac1c$. Therefore, 
    \[
        \textup{Var}\big[ S / c\big] = \sum_{j} \textup{Var}\big[ {X_j}/c\big] \leq /{c} \sum_{j } \EE\big[{X_j}/c\big] \leq \frac1c , 
    \] 
    where we used the independence of the~$X_j$ and the fact that~$\sum_{j } \EE[X_j] \leq c$. We fix some~$t \geq 1$. Bernstein's inequality (\Cref{lem_bernstein}) guarantees 
    \[
        \PP\bigg[\frac S c \geq 2 + t\bigg] \leq 
        \exp \bigg( - \frac{t^2}{2/c + 2t/3} \bigg) \leq \exp \bigg( - c \frac{t^2}{2 + 2t/3} \bigg) \leq \exp\bigg( -\frac{c}{3} t  \bigg),
    \]
    where we used that~$c \geq 1$. 
    Using a union bound, we obtain 
    \begin{align*}
        \PP \Big[\max_{1 \leq i \leq m} S_i /c_i \geq 2 + t \Big] & \leq \sum_{i=1}^{m} \exp\bigg( -\frac{c_i}{3} t  \bigg) \\
        & \leq \sum_{i=1}^{m} \exp\bigg( - (3/2)^{m-i} c_{m} \frac{t}{3} \bigg) \\ 
        & \leq \sum_{i=1}^{m} \exp\bigg( - (3/2)^{m-i}\frac{t}{3} \bigg) \\ 
        & \leq \sum_{i=1}^{m} \bigg(\frac23\bigg)^{m-i} e^{-t/3} \\
        & \leq 3 e^{-t/3},
    \end{align*}
    where we used~$c_i \geq \frac32 c_{i+1}$ in the second inequality and~$c_{m} \geq 1$ in the third inequality. For the fourth inequality, we fix~$t \geq 3$ and use that, for~$s \geq 1$ and~$x \geq 0$, we have~$\exp\big( - s \big(\frac32\big)^x \big) \leq \big(\frac23\big)^x e^{-s}$. 
    
    Hence, we conclude the proof with \[
        \EE\big[\max_{1 \leq i \leq m} S_i/c_i \big] \leq (2 + 6) + \int_{5}^\infty \PP\big[ \max_{1 \leq i \leq m} S_i/c_i > 2 + t \big] dt \leq 8 + 3 \int_{5}^\infty e^{-t/5} dt = 8 + 15/e = O(1). 
    \]
\end{proof}

\section{Feasibility of \eqref{eq_lpstochconfig}}\label{app_config}


\lemlpfeasible*

\begin{proof}
    Consider the adaptive policy guaranteed by \Cref{thm_adap_restart}, and let the events $j \rightarrow c$ be with respect to this policy. 
    We consider the natural setting of the $y$-variables given this policy: for all~$j$ and~$c \in [q_j]$, we take $y_{cj} = \mathbb{P}(j \rightarrow c)$. It is clear that $\sum_{c=1}^{q_j} y_{cj} = 1$ for all~$j$ and $0 \leq y_{cj} \leq 1$ for all~$j$ and for all~$c\in [q_j]$. Moreover, as by \Cref{thm_adap_restart} the policy does not select a configuration~$c$ with~$\EE\big[ \max_i X_i(c) \big] > \tau$, the pruning constraints~$y_{cj} = 0$ if $\EE\big[ \max_i X_i(c) \big] > \tau$ are also satisfied.
    
    It remains to verify the exceptional and truncated load constraints. For the exceptional constraint, we observe
    \[\sum_{j=1}^n\sum_{c=1}^{q_j} \mathbb{E}\Big[\max_i X_{ij}^E(c)\Big] \cdot y_{cj} = \sum_{j=1}^n\sum_{c=1}^{q_j} \mathbb{E}\Big[\max_i X_{ij}^E(c) \cdot \one_{j \rightarrow c}\Big] \leq 2 \cdot \mathbb{E}[\Opt] \leq \gamma,\]
    where in the first step we use the fact that the decision to choose configuration $c$ for request $j$ is independent of its realization, and in the second we use the properties of the policy. Similarly, for the truncated constraint for each $i$,
    \[\sum_{j=1}^n\sum_{c=1}^{q_j} \mathbb{E}\big[X_{ij}^T(c)\big] \cdot y_{cj} = \sum_{j=1}^n\sum_{c=1}^{q_j} \mathbb{E}\big[X_{ij}^T(c) \cdot \one_{j \rightarrow c}\big] \leq 2 \cdot \mathbb{E}[\Opt] \leq \tau.\]
\end{proof}

\section{Machine smoothing analysis}\label{sec_appendix_related}

\lemrelsmooth*
\begin{proof}
    It is clear that the algorithm is efficient and outputs $\mathcal{I}_s$ satisfying~\ref{smoothed_prop_speeds} and~\ref{smoothed_prop_groups}. It remains to show~\ref{smoothed_prop_cost}. To do so, we show that each step increases \Opt{} by only a constant factor.
    
    Recall that \Opt{} is an adaptive policy, so when \Opt{} decides to schedule job~$j$ on machine~$i$, it immediately learns the realized value of~$X_{ij}$, or equivalently~$X_j = s_i \cdot X_{ij}$ on related machines. Hence, we may assume that the next scheduling decision of \Opt{} is completely determined by the previously realized~$X_j$s. In particular, if \Opt{} decides to schedule~$j$ on machine~$i$, then we can modify \Opt{} by scheduling~$j$ on some other machine~$i'$ and leaving the subsequent decisions (which in general depend on~$X_j$) unchanged. Similarly, if we modify the speed of machine~$i$, then this does not affect subsequent scheduling decisions. 
    
    \begin{enumerate}
        \item[\ref{alg_ms_delete_slow}]  Let $\Opt = \Opt(\mathcal{I})$ denote the initial optimal policy. Consider the following adaptive policy for the instance after step \ref{alg_ms_delete_slow}: If \Opt{} schedules~$j$ on~$i$ such that~$i$ is not deleted, then we also schedule~$j$ on~$i$. Otherwise, \Opt{} schedules~$j$ on~$i$ such that~$i$ is deleted for being too slow. Then we schedule~$j$ on the fastest machine. 
        
        For every realization of job sizes, the modified policy only increases the load of the fastest machine. We delete at most~$m-1$ machines each having speed at most~$\frac{1}{m}$. Thus, we schedule all jobs assigned to these machines on a machine that is at least~$m$ times faster. The increase in load on the fastest machine is thus at most~$(m-1) \cdot \frac{\Opt}{m} \leq \Opt$.
        
        After deleting all slow machines, all machine speeds are in~$\big(\frac1m, 1 \big]$.
        \item[\ref{alg_ms_round_down}] We decrease the speed of each machine by at most a factor~$2$, so the makespan of the optimal policy increases by at most a factor~$2$. 
        
        Further, after rounding down the machine speeds, there are at most~$\lceil \log m \rceil$ distinct speeds and thus groups. 
        \item[\ref{alg_ms_delete_small}] First, we note that we keep at least one group, namely the fastest one. Consider any group~$k$ that is deleted, and let~$k' > k$ be the fastest subsequent group that is kept. Because we delete all groups between~$k$ and~$k'$, we have~$m_k < \big(\frac{3}{2}\big)^{k' - k} \cdot m_{k'}$. Further, because all groups have distinct speeds that differ by at least a factor~$2$, we also have~$s_{k'} \geq 2^{k' - k} \cdot s_k$.
        
        Let \Opt{} be the optimal policy after step \ref{alg_ms_round_down}. We re-assign the jobs that \Opt{} schedules on group~$k$ to group~$k'$ as follows. Because~$m_k < \big(\frac{3}{2}\big)^{k' - k} \cdot m_{k'}$, we fix a mapping from the machines in group~$k$ to those of~$k'$ such that each machine in~$k'$ is mapped to by at most~$\big(\frac{3}{2}\big)^{k' - k}$ machines in~$k$. Then, when \Opt{} schedules a job on a machine in group~$k$, we instead schedule it on the machine it maps to in group~$k'$. This completes the description of our modified policy.
        
        To bound the makespan, consider any machine~$i$ in a kept group~$k'$. We upper-bound the increase in load on~$i$ due to re-assignments from slower deleted groups. For any deleted group~$k < k'$, at most~$\big(\frac{3}{2}\big)^{k' - k}$ machines from group~$k$ map to~$i$. Each such machine in group~$k$ under policy \Opt{} has load at most \Opt{}. However, recall that~$i$ is at least a~$2^{k'-k}$-factor faster than any machine in group~$k$, so the increase in load on machine~$i$ due to deleted machines from group~$k < k'$ is at most~$\big(\frac{3}{2}\big)^{k'-k} \cdot 2^{-(k' -k)} \cdot \Opt = \big(\frac{3}{4}\big)^{k' -k} \cdot \Opt$. Summing over all~$k < k'$, the total increase in load on a machine in group~$k'$ is at most~$\sum_{k < k'} \big(\frac{3}{4}\big)^{k' -k} \cdot \Opt = O(\Opt)$.\endproof
    \end{enumerate}
\end{proof}

\section{Clairvoyance Gap for Load Balancing on Related Machines}\label{appendix_nonclairvoant_related}


In this section, we quantify the power and limitation of non-clairvoyant algorithms for load balancing on related machines. Recall that a non-clairvoyant algorithm has no prior knowledge of the job sizes and must decide where to schedule a job using only the information about current machine loads and speeds. After an algorithm's decision upon assigning a job, an adversary chooses the realized size of the job, which the algorithm then observes for subsequent decisions. We refer to a non-clairvoyant {\em online} algorithm as a non-clairvoyant algorithm that learns about the existence of jobs (of unknown size) online one by one.

Our goal is to approximate the optimal makespan achievable by a clairvoyant offline algorithm, which knows all jobs and sizes in advance, by a non-clairvoyant (online) algorithm. We define the {\em clairvoyance gap} as the maximum ratio between the makespan of an optimal non-clairvoyant algorithm and the optimal makespan achievable by a clairvoyant offline algorithm.

We show the following result.
\begin{theorem}
	The clairvoyance gap for related machine scheduling is $\Theta(\sqrt{m})$.
\end{theorem}

To show this result, we first give a lower bound on the clairvoyance gap and then complement it by an upper bound achieved by a non-clairvoyant online list scheduling algorithm.

\begin{lem}\label{lem_nonclair_lower}
    The clairvoyance gap for scheduling on related machines is $\Omega(\sqrt{m})$.
\end{lem}
\begin{proof}
    Consider an instance with~$m$ machines, one fast machine with speed~$1$ and~$m-1$ slow machines with speed~$\frac1{\sqrt m}$. There are~$m$ jobs, one big job with size~$1$ and~$m-1$ small jobs with size $\frac{1}{\sqrt{m}}$.    There is a schedule of makespan $1$, which is achieved by assigning the big job to the fast machine and one small job per slow machine.

    

	We show that any non-clairvoyant algorithm has makespan $\Omega(\sqrt{m})$ which implies the lemma.
    To that end, we define the following adversarial strategy: the adversary reveals only small jobs (unless it runs out of small jobs, in which case the final job is big) until the first time that the algorithm decides to use a slow machine. Then the adversary makes this job big.   
    If an algorithm only uses the fast machine, then it has makespan $1 + (m-1) \cdot \frac{1}{\sqrt{m}} = \Omega(\sqrt{m})$. Otherwise, the algorithm must use a slow machine, which receives the big job from the adversary, which also gives makespan $\sqrt{m}$. Hence, any non-clairvoyant algorithm incurs a makespan $\Omega(\sqrt{m})$.
\end{proof}


In the next lemma, we show an upper bound on the clairvoyance gap. We refer to {\em list scheduling} as the algorithm that assigns the next job to the currently least-loaded machine. Here, the load of a machine is the total processing time of jobs that have been assigned to the machine. We use list scheduling only on a subset of  machines that are ``fast enough'' and leave the remaining machines idle. 
More precisely, we list schedule the jobs in the order of their arrival on the machines with speeds within a factor~$\frac1{\sqrt{m}}$ of the fastest.

\begin{lem}\label{lem_nonclair_upper}
    The clairvoyance gap for related machine scheduling is~$O(\sqrt m)$. 
\end{lem}

\begin{proof}
    For convenience, we re-scale the instance such that the speed of the fastest machine is~$1$. Hence, our algorithm uses all machines~$i$ with~$s_i \in \big[ \frac1{\sqrt m},1\big]$; we call such machines \emph{fast} and the remaining ones \emph{slow}. 
    Clearly, list scheduling on the fast machines is both, online and non-clairvoyant. It remains to bound the competitive ratio. To this end, we observe 
    \[
        C_{\max} \leq \frac{\sum_{j \in J} p_j}{\sum_{i: s_i \geq 1/\sqrt{m}} s_i} + \frac{\max_{j \in J} p_j}{\min_{i: s_i \geq 1/\sqrt m} s_i},  
    \]
    where the first term bounds the starting time and the second term bounds the processing time of any job. We bound both terms separately by comparing to the optimal solution \Opt, where \Opt{} denotes the optimal schedule as well as its makespan. Clearly,~$\max_{j \in J} p_j \leq \Opt$, which implies that the second term is at most~$\sqrt m \cdot \Opt$. 
    
    For the first term, note that the total size of jobs that \Opt{} schedules on fast machines is at most~$m'\cdot \Opt$, where~$m'$ is the number of fast machines. The total size of jobs that \Opt{} schedules on slow machines is at most~$m \cdot \frac1{\sqrt m} \cdot \Opt = \sqrt m \cdot \Opt$ as there are at most~$m$ slow machines. Further, the total speed of the fast machines is at least~$1$ (because the fastest machine has speed~1) and also at least~$m' \frac1{\sqrt m}$. Therefore,
    \[
        \frac{\sum_{j \in J} p_j}{\sum_{i: s_i \geq 1/\sqrt{m}} s_i} \leq \frac{m' \cdot \Opt + \sqrt m \cdot \Opt}{\max\{ 1, m'/\sqrt m \} } \leq \bigg(\frac{m'}{m'/\sqrt m} + \frac{\sqrt m}{1} \bigg) \cdot \Opt = O(\sqrt m) \cdot \Opt,
    \]
    concluding the proof.
\end{proof}

{\small
\bibliographystyle{alpha}
\bibliography{ref}

\newcommand{\etalchar}[1]{$^{#1}$}
\begin{thebibliography}{GKMR11}

\bibitem[AAF{\etalchar{+}}97]{AspnesAFPW97}
James Aspnes, Yossi Azar, Amos Fiat, Serge~A. Plotkin, and Orli Waarts.
\newblock On-line routing of virtual circuits with applications to load
  balancing and machine scheduling.
\newblock {\em J. {ACM}}, 44(3):486--504, 1997.

\bibitem[AD15]{AD15}
Shipra Agrawal and Nikhil~R. Devanur.
\newblock Fast algorithms for online stochastic convex programming.
\newblock In {\em Proceedings of {SODA}}, pages 1405--1424, 2015.

\bibitem[ANR95]{DBLP:journals/jal/AzarNR95}
Yossi Azar, Joseph Naor, and Raphael Rom.
\newblock The competitiveness of on-line assignments.
\newblock {\em J. Algorithms}, 18(2):221--237, 1995.

\bibitem[AWY14]{AWY14}
Shipra Agrawal, Zizhuo Wang, and Yinyu Ye.
\newblock A dynamic near-optimal algorithm for online linear programming.
\newblock {\em Operations Research}, 62(4):876--890, 2014.

\bibitem[Aza96]{Azar96}
Yossi Azar.
\newblock On-line load balancing.
\newblock In {\em Online Algorithms}, volume 1442 of {\em Lecture Notes in
  Computer Science}, pages 178--195. Springer, 1996.

\bibitem[BCK00]{BermanCK00}
Piotr Berman, Moses Charikar, and Marek Karpinski.
\newblock On-line load balancing for related machines.
\newblock {\em J. Algorithms}, 35(1):108--121, 2000.

\bibitem[Bea55]{beale55}
E.M.L. Beale.
\newblock On minimizing a convex function subject to linear inequalities.
\newblock {\em Journal of the Royal Statistical Society. Series B.
  Methodological}, 17:173--184; discussion, 194--203, 1955.

\bibitem[Bel58]{Bellman58}
Richard Bellman.
\newblock On a routing problem.
\newblock {\em Quarterly of Applied Mathematics}, 16:87--90, 1958.

\bibitem[Ben62]{Bennett62}
George Bennett.
\newblock Probability inequalities for the sum of independent random variables.
\newblock {\em Journal of the American Statistical Association},
  57(297):33--45, 1962.

\bibitem[BGK11]{BhalgatGK11}
Anand Bhalgat, Ashish Goel, and Sanjeev Khanna.
\newblock Improved approximation results for stochastic knapsack problems.
\newblock In {\em Proceedings of {SODA}}, pages 1647--1665. {SIAM}, 2011.

\bibitem[CCP05]{CharikarCP2005}
Moses Charikar, Chandra Chekuri, and Martin P{\'a}l.
\newblock Sampling bounds for stochastic optimization.
\newblock In {\em Proceedings of {APPROX}}, volume 3624 of {\em LNCS}, pages
  257--269, 2005.

\bibitem[CGKT07]{ChuzhoyGKT07}
Julia Chuzhoy, Venkatesan Guruswami, Sanjeev Khanna, and Kunal Talwar.
\newblock Hardness of routing with congestion in directed graphs.
\newblock In {\em {STOC}}, pages 165--178. {ACM}, 2007.

\bibitem[CR06]{ChawlaR06}
Shuchi Chawla and Tim Roughgarden.
\newblock Single-source stochastic routing.
\newblock In {\em Proceedings of APPROX}, pages 82--94. Springer, 2006.

\bibitem[Dan55]{dantzig55}
G.B. Dantzig.
\newblock {Linear programming under uncertainty}.
\newblock {\em Management Science}, 1:197--206, 1955.

\bibitem[DGV08]{deanGV08}
B.C. Dean, M.X. Goemans, and J.~Vondr{\'a}k.
\newblock Approximating the stochastic knapsack problem: The benefit of
  adaptivity.
\newblock {\em Mathematics of Operations Research}, 33(4):945--964, 2008.

\bibitem[DP09]{DubhashiP2009}
Devdatt~P. Dubhashi and Alessandro Panconesi.
\newblock {\em Concentration of Measure for the Analysis of Randomized
  Algorithms}.
\newblock Cambridge University Press, 2009.

\bibitem[DST03]{DyeST2003}
Shane Dye, Leen Stougie, and Asgeir Tomasgard.
\newblock The stochastic single resource service-provision problem.
\newblock {\em Naval Res. Logist.}, 50(8):869--887, 2003.

\bibitem[GK17]{GuptaK17}
Anupam Gupta and Archit Karandikar.
\newblock Stochastic unsplittable flows.
\newblock In {\em {APPROX-RANDOM}}, volume~81 of {\em LIPIcs}, pages 7:1--7:19.
  Schloss Dagstuhl - Leibniz-Zentrum f{\"{u}}r Informatik, 2017.

\bibitem[GKMR11]{GuptaKMR11}
Anupam Gupta, Ravishankar Krishnaswamy, Marco Molinaro, and R.~Ravi.
\newblock Approximation algorithms for correlated knapsacks and non-martingale
  bandits.
\newblock In Rafail Ostrovsky, editor, {\em {IEEE} 52nd Annual Symposium on
  Foundations of Computer Science, {FOCS} 2011, Palm Springs, CA, USA, October
  22-25, 2011}, pages 827--836. {IEEE} Computer Society, 2011.

\bibitem[GKNS21]{DBLP:journals/mor/Gupta0NS21}
Anupam Gupta, Amit Kumar, Viswanath Nagarajan, and Xiangkun Shen.
\newblock Stochastic load balancing on unrelated machines.
\newblock {\em Math. Oper. Res.}, 46(1):115--133, 2021.

\bibitem[GKNS22]{DBLP:journals/mp/GuptaKNS22}
Anupam Gupta, Amit Kumar, Viswanath Nagarajan, and Xiangkun Shen.
\newblock Stochastic makespan minimization in structured set systems.
\newblock {\em Math. Program.}, 192(1):597--630, 2022.

\bibitem[GM16]{GM-MOR16}
Anupam Gupta and Marco Molinaro.
\newblock How the experts algorithm can help solve lps online.
\newblock {\em Math. Oper. Res.}, 41(4):1404--1431, 2016.

\bibitem[GMUX20]{GuptaMUX20}
Varun Gupta, Benjamin Moseley, Marc Uetz, and Qiaomin Xie.
\newblock Greed works - online algorithms for unrelated machine stochastic
  scheduling.
\newblock {\em Math. Oper. Res.}, 45(2):497--516, 2020.

\bibitem[GPRS11]{GuptaPRS2011}
Anupam Gupta, Martin P{\'a}l, R.~Ravi, and Amitabh Sinha.
\newblock Sampling and cost-sharing: {Approximation} algorithms for stochastic
  optimization problems.
\newblock {\em SIAM J. Comput.}, 40(5):1361--1401, 2011.

\bibitem[Gra69]{DBLP:journals/siamam/Graham69}
Ronald~L. Graham.
\newblock Bounds on multiprocessing timing anomalies.
\newblock {\em {SIAM} Journal of Applied Mathematics}, 17(2):416--429, 1969.

\bibitem[IKKP19]{ImKKP19}
Sungjin Im, Nathaniel Kell, Janardhan Kulkarni, and Debmalya Panigrahi.
\newblock Tight bounds for online vector scheduling.
\newblock {\em {SIAM} J. Comput.}, 48(1):93--121, 2019.

\bibitem[IKPS18]{ImKPS18}
Sungjin Im, Nathaniel Kell, Debmalya Panigrahi, and Maryam Shadloo.
\newblock Online load balancing on related machines.
\newblock In {\em {STOC}}, pages 30--43. {ACM}, 2018.

\bibitem[IMP15]{ImMP15}
Sungjin Im, Benjamin Moseley, and Kirk Pruhs.
\newblock Stochastic scheduling of heavy-tailed jobs.
\newblock In Ernst~W. Mayr and Nicolas Ollinger, editors, {\em Proceedings of
  32nd International Symposium on Theoretical Aspects of Computer Science
  ({STACS})}, volume~30, pages 474--486, 2015.

\bibitem[KRT00]{DBLP:journals/siamcomp/KleinbergRT00}
Jon~M. Kleinberg, Yuval Rabani, and {\'{E}}va Tardos.
\newblock Allocating bandwidth for bursty connections.
\newblock {\em {SIAM} J. Comput.}, 30(1):191--217, 2000.

\bibitem[Leo96]{Leonardi96}
Stefano Leonardi.
\newblock On-line network routing.
\newblock In {\em Online Algorithms}, volume 1442 of {\em Lecture Notes in
  Computer Science}, pages 242--267. Springer, 1996.

\bibitem[LRS98]{LeightonRS98}
Tom Leighton, Satish Rao, and Aravind Srinivasan.
\newblock Multicommodity flow and circuit switching.
\newblock In {\em {HICSS} {(7)}}, pages 459--465. {IEEE} Computer Society,
  1998.

\bibitem[LST90]{DBLP:journals/mp/LenstraST90}
Jan~Karel Lenstra, David~B. Shmoys, and {\'{E}}va Tardos.
\newblock Approximation algorithms for scheduling unrelated parallel machines.
\newblock {\em Math. Program.}, 46:259--271, 1990.

\bibitem[Ma18]{Ma18}
Will Ma.
\newblock Improvements and generalizations of stochastic knapsack and markovian
  bandits approximation algorithms.
\newblock {\em Math. Oper. Res.}, 43(3):789--812, 2018.

\bibitem[Mol19]{Molinaro19}
Marco Molinaro.
\newblock Stochastic {\(\mathscr{l}\)}p load balancing and moment problems via
  the l-function method.
\newblock In {\em {SODA}}, pages 343--354. {SIAM}, 2019.

\bibitem[MSU99a]{DBLP:journals/jacm/MohringSU99}
Rolf~H. M{\"{o}}hring, Andreas~S. Schulz, and Marc Uetz.
\newblock Approximation in stochastic scheduling: the power of lp-based
  priority policies.
\newblock {\em J. {ACM}}, 46(6):924--942, 1999.

\bibitem[MSU99b]{MohringSU99}
Rolf~H. M{\"{o}}hring, Andreas~S. Schulz, and Marc Uetz.
\newblock Approximation in stochastic scheduling: the power of lp-based
  priority policies.
\newblock {\em J. {ACM}}, 46(6):924--942, 1999.

\bibitem[MUV06]{MegowUV06}
Nicole Megow, Marc Uetz, and Tjark Vredeveld.
\newblock Models and algorithms for stochastic online scheduling.
\newblock {\em Math. Oper. Res.}, 31(3):513--525, 2006.

\bibitem[RT87]{RaghavanT87}
Prabhakar Raghavan and Clark~D. Thompson.
\newblock Randomized rounding: a technique for provably good algorithms and
  algorithmic proofs.
\newblock {\em Comb.}, 7(4):365--374, 1987.

\bibitem[Sch08]{schulz08}
Andreas~S. Schulz.
\newblock Stochastic online scheduling revisited.
\newblock In Boting Yang, Ding{-}Zhu Du, and Cao~An Wang, editors, {\em
  Combinatorial Optimization and Applications, Second International Conference,
  {COCOA} 2008, St. John's, NL, Canada, August 21-24, 2008. Proceedings},
  volume 5165 of {\em Lecture Notes in Computer Science}, pages 448--457.
  Springer, 2008.

\bibitem[SS12]{SwamyS2012}
Chaitanya Swamy and David~B. Shmoys.
\newblock Sampling-based approximation algorithms for multistage stochastic
  optimization.
\newblock {\em SIAM J. Comput.}, 41(4):975--1004, 2012.

\bibitem[SS21]{SagnolW21}
Guillaume Sagnol and Daniel {Schmidt genannt Waldschmidt}.
\newblock Restricted adaptivity in stochastic scheduling.
\newblock In {\em {ESA}}, volume 204 of {\em LIPIcs}, pages 79:1--79:14.
  Schloss Dagstuhl - Leibniz-Zentrum f{\"{u}}r Informatik, 2021.

\bibitem[SSU16]{SkutellaSU16}
Martin Skutella, Maxim Sviridenko, and Marc Uetz.
\newblock Unrelated machine scheduling with stochastic processing times.
\newblock {\em Math. Oper. Res.}, 41(3):851--864, 2016.

\bibitem[ST93]{DBLP:journals/mp/ShmoysT93}
David~B. Shmoys and {\'{E}}va Tardos.
\newblock An approximation algorithm for the generalized assignment problem.
\newblock {\em Math. Program.}, 62:461--474, 1993.

\bibitem[SU01]{SkutellaU01}
Martin Skutella and Marc Uetz.
\newblock Scheduling precedence-constrained jobs with stochastic processing
  times on parallel machines.
\newblock In {\em {SODA}}, pages 589--590. {ACM/SIAM}, 2001.

\end{thebibliography}
}

\end{document}